\documentclass[a4paper, amsfonts, amssymb, amsmath, reprint, showkeys, nofootinbib, twoside, aip, floatfix]{revtex4-2}
\usepackage[english]{babel}
\usepackage[utf8]{inputenc}
\usepackage[colorinlistoftodos, color=green40, prependcaption]{todonotes}
\usepackage{amsthm}
\usepackage{mathtools}
\usepackage{xcolor}
\usepackage{graphicx}
\usepackage[left=23mm,right=13mm,top=20mm,bottom=20mm,columnsep=15pt]{geometry} 
\usepackage{adjustbox}
\usepackage{placeins}
\usepackage[T1]{fontenc}
\usepackage{lipsum}
\usepackage{csquotes}
\usepackage{siunitx}
\usepackage[caption=false]{subfig}

\makeatletter 
\renewcommand\onecolumngrid{
\do@columngrid{one}{\@ne}%
\def\set@footnotewidth{\onecolumngrid}
\def\footnoterule{\kern-6pt\hrule width 1.5in\kern6pt}%
}
 
\renewcommand\twocolumngrid{
        \def\footnoterule{
        \dimen@\skip\footins\divide\dimen@\thr@@
        \kern-\dimen@\hrule width.5in\kern\dimen@}
        \do@columngrid{mlt}{\tw@}
}%
 
\makeatother

\usepackage{fancyhdr}

\makeatletter
 
\newcommand{\@aionnum}{xxxxxx}
 
\fancypagestyle{plain}{%
\fancyhf{} 
\fancyhead[R]{AION-REPORT/\@aionnum}
}
 
\newcommand{\aionnum}[1]{%
    \def\@aionnum{#1}%
}

\makeatother

\usepackage[pdftex, pdftitle={Article}, pdfauthor={Author}]{hyperref} 
\usepackage[T1]{fontenc}
\usepackage{wasysym}
\usepackage{cancel}
\usepackage{multirow}
\usepackage[splitrule,bottom,hang]{footmisc}

\newcommand{\sreights}{$^{88}$Sr}
\newcommand{\transition}{$^1$S$_0 \rightarrow ^1$P$_1$}
\newcommand{\toven}{$T_\text{oven}$}
\newcommand{\mfp}{$\lambda_\text{mf}$}

\begin{document}

\title{A high-flux atomic strontium oven with light-driven flux modulation}
\aionnum{2026-02}

\author{Kenneth M. Hughes}
    \email{kenneth.hughes@physics.ox.ac.uk}
    \affiliation{Department of Physics, University of Oxford, Parks Road, Oxford, OX1 3PU, UK}
\author{Jesse S. Schelfhout}
    \affiliation{Department of Physics, University of Oxford, Parks Road, Oxford, OX1 3PU, UK}
\author{Charu Mishra}
    \altaffiliation{Present address: Department of Physics and Astronomy, University of Sussex, Brighton, BN1 9QH}
    \affiliation{Department of Physics, University of Oxford, Parks Road, Oxford, OX1 3PU, UK}
\author{Timothy Leese}
    \affiliation{Department of Physics, University of Oxford, Parks Road, Oxford, OX1 3PU, UK}
\author{Elliot Bentine}
    \affiliation{Department of Physics, University of Oxford, Parks Road, Oxford, OX1 3PU, UK}
\author{Christopher J. Foot}
    \email{christopher.foot@physics.ox.ac.uk}
    \affiliation{Department of Physics, University of Oxford, Parks Road, Oxford, OX1 3PU, UK}



\begin{abstract}
A high-flux source of strontium atoms is required for cold atom quantum technology applications. We present a re-entrant oven design that avoids the need for any vacuum feed-throughs and has an inherent temperature gradient to guard against clogging of the nozzle. The nozzle is fabricated by micro-machining of fused silica using selective laser etching; this specialised technique is capable of making many thousands of fine microchannels and is suitable for batch production. Operating with only electrical heating, using <\qty{20}{\watt} of electrical power, a total flux of \num{8(1)e14} atoms/s is achieved at an oven temperature of \qty{475}{\degree}C, of which we estimate \num{1.8(2)e13} atoms/s could be captured. A heated in-vacuum sapphire window grants optical access directly opposite the oven, and can be cleared of metallization without breaking vacuum. We used this optical access to modulate the flux of the atomic beam by direct illumination of the nozzle and the strontium metal with high-power laser light. Heating by laser light increased the useful flux by a factor of up to 16(3) on a timescale of \qty{40}{\second}, and a factor of 2.5(5) on a timescale of \qty{1}{\second}. This flux modulation serves to increase the operating lifetime of the oven. We report experimental measurements of the performance of the oven in long-term operation over many months.
\end{abstract}


\maketitle
\thispagestyle{plain}

\onecolumngrid


\section{Introduction}
\label{sec-intro}

An efficient and high-flux atomic beam is a critical starting point of laser cooling and trapping experiments, providing the basis of many experiments in atomic physics \cite{oxley2016precision,nosske2017two,bowden2019realize,graham2017midbandgravitationalwavedetection,Alonso2022,stellmer2013production,escudero2021steady} and for enabling the development of advanced quantum technologies. For alkaline earth atoms such as strontium, high temperatures are required to achieve a vapour pressure to produce an atomic beam for cold atom experiments. This can be achieved with the use of an oven.

Various types of atomic oven have been demonstrated, optimized for a range of applications such as a high flux, a highly collimated atomic beam, or a compact source. Atomic ovens are typically designed to operate as effusive sources
\cite{feng2024high,schioppo2012compact,hahn2022comparative}, and commonly use a multichannel capillary array to collimate the output atomic beam, reducing the flux of atoms at large angles that cannot be trapped \cite{ross1995high, senaratne2015effusive,yang2015high,bowden2016adaptable,song2016cost,lebedev2017note,cooper2018collimated,li2020robust}. Stacking capillary tubes is straightforward, but not easily scalable to large channel numbers nor for batch production of sources.

In this paper, we present a compact, re-entrant oven design with a laser etched glass nozzle and a heated sapphire window through which we modulate the atom flux with high-power laser light. This approach achieves a high flux, is efficient in usage of strontium, and can be manufactured at scale. Nozzles with over \num{100000} holes have been fabricated by this method but not yet tested; the proof-of-principle measurements reported here indicate the utility of this approach and that there is a large scope for developing it further. 

Previous work on atom sources that use laser light to produce the atomic vapour ranges from light-induced atomic desorption, for example using ultraviolet light for alkali metals such as Rb \cite{anderson2001loading}, to laser-induced thermal ablation, which has been used for Sr \cite{hsu2022laser}. Similar experiments have used UV light on SrO to produce Sr \cite{kock2016laser} and on Yb$_2$O$_3$ to produce Yb \cite{yasuda2017laser}. In almost all cases, the light is focussed down to sub-millimetre beam waists, and in used to produce a vapour in a glass cell rather than for applications that require an atomic beam. The use of light to produce a collimated beam of Sr via optical heating has been demonstrated for systems requiring low fluxes (\num{\sim e7} atoms/s) \cite{gao2021optically}.
%
%
%
We demonstrate here a system that combines electrical and optical heating for a high flux of Sr atoms (up to >\num{e14} atoms/s) in a collimated beam.

Modulation of the flux from the oven by using a pulse of high-power laser light to briefly increase the flux allows the system to be run at a lower base temperature, while also reducing the amount of Sr deposited on the chamber walls; the build-up of dendritic whiskers can interfere with optical beam paths. This extends the lifetime of the oven, defined as the time taken for the supply of Sr metal in the oven to be fully depleted. Using a pulse duration of \qty{1}{\second}, we measured the flux that could feasibly be cooled and trapped to increase by a factor of up to 2.5(5) when operating at a total flux of $\mathcal{O}(\num{e14})$ atoms/s, dropping to 1.5(3) at $\mathcal{O}(\num{e15})$ atoms/s. The absolute flux and the fractional increase are both considerably higher for longer heating light pulses, however these are less well-suited for use in cold atom experiments. The results reported in this paper demonstrate a robust method for collimation and modulation of an atomic beam of strontium suitable for long-term operation. However, we have not reached the limits of this approach and even higher degrees of collimation and atomic fluxes could be obtained with further development. 
%

\section{Oven Design}
\label{sec-oven}

\begin{figure}[!t]
     \centering
     \subfloat[\label{fig-oven}]{%
         \def\svgwidth{0.24\columnwidth}
         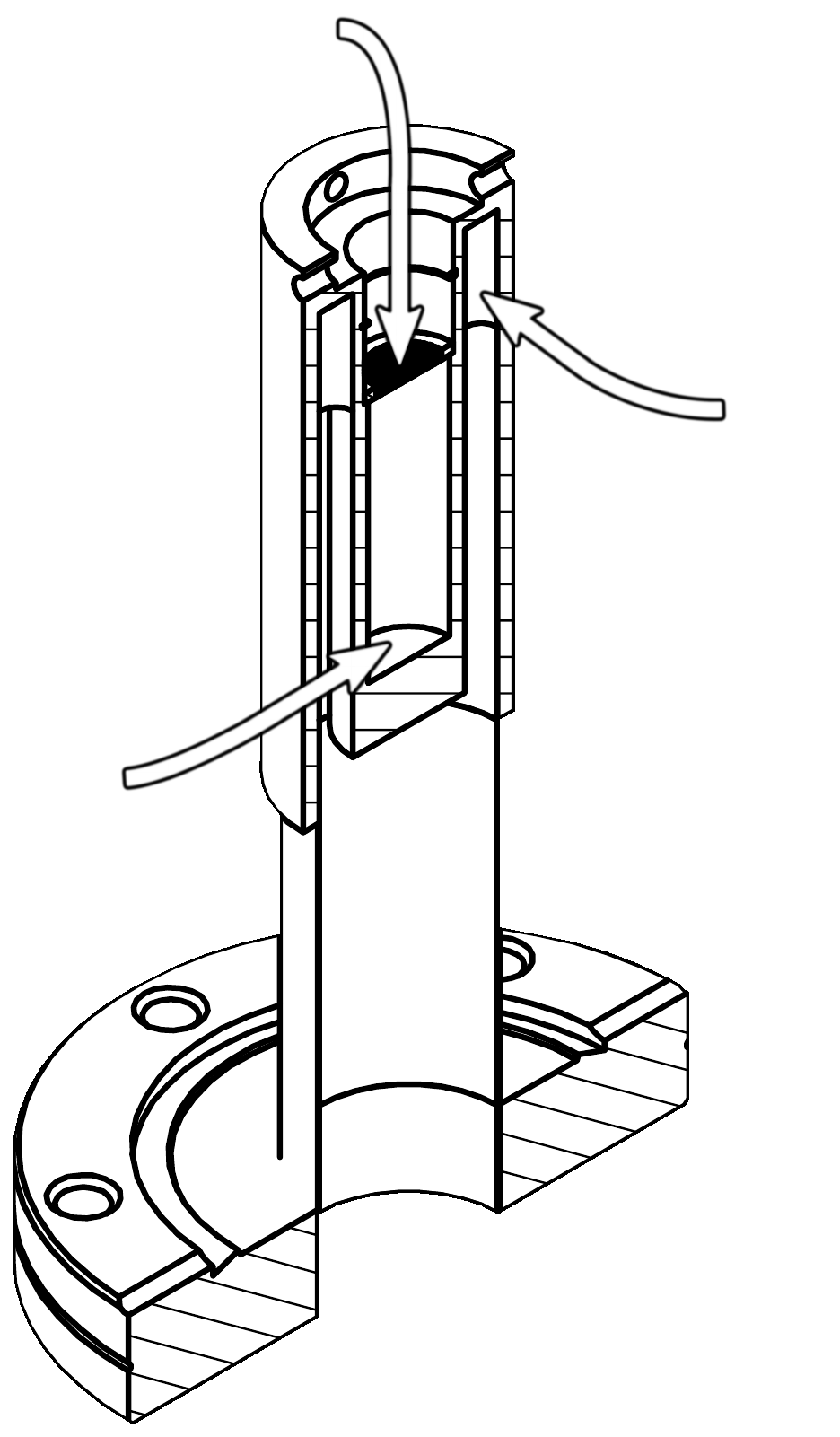%
         }
     \hspace{5mm}
     \subfloat[\label{fig-sapphirewindow}]{%
         \def\svgwidth{0.24\columnwidth}
\begingroup%
  \makeatletter%
  \providecommand\color[2][]{%
    \errmessage{(Inkscape) Color is used for the text in Inkscape, but the package 'color.sty' is not loaded}%
    \renewcommand\color[2][]{}%
  }%
  \providecommand\transparent[1]{%
    \errmessage{(Inkscape) Transparency is used (non-zero) for the text in Inkscape, but the package 'transparent.sty' is not loaded}%
    \renewcommand\transparent[1]{}%
  }%
  \providecommand\rotatebox[2]{#2}%
  \newcommand*\fsize{\dimexpr\f@size pt\relax}%
  \newcommand*\lineheight[1]{\fontsize{\fsize}{#1\fsize}\selectfont}%
  \ifx\svgwidth\undefined%
    \setlength{\unitlength}{253.9394989bp}%
    \ifx\svgscale\undefined%
      \relax%
    \else%
      \setlength{\unitlength}{\unitlength * \real{\svgscale}}%
    \fi%
  \else%
    \setlength{\unitlength}{\svgwidth}%
  \fi%
  \global\let\svgwidth\undefined%
  \global\let\svgscale\undefined%
  \makeatother%
  \begin{picture}(1,1.62094537)%
    \lineheight{1}%
    \setlength\tabcolsep{0pt}%
    \put(0,0){\includegraphics[width=\unitlength,page=1]{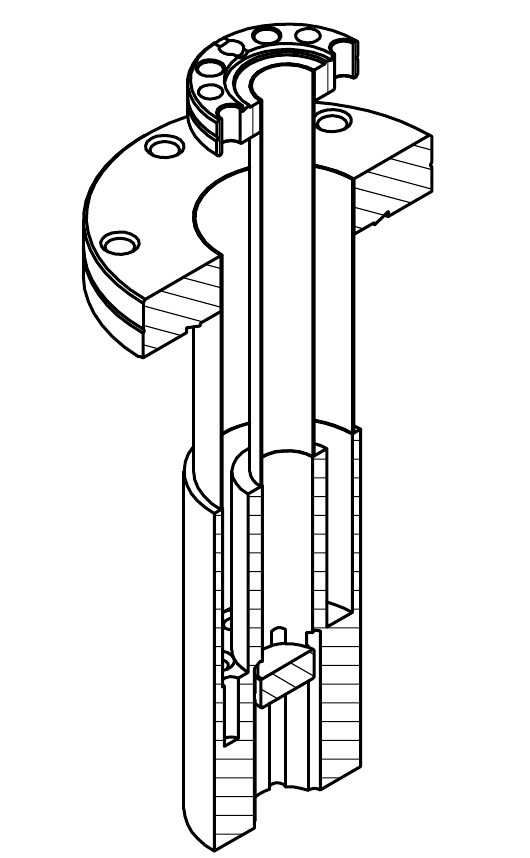}}%
    \put(0.04979372,1.5329009){\color[rgb]{0,0,0}\makebox(0,0)[lt]{\lineheight{1.25}\smash{\begin{tabular}[t]{l}VP\end{tabular}}}}%
    \put(-0.00530699,0.31871173){\color[rgb]{0,0,0}\makebox(0,0)[lt]{\lineheight{1.25}\smash{\begin{tabular}[t]{l}CH\end{tabular}}}}%
    \put(0.84618942,0.19255205){\color[rgb]{0,0,0}\makebox(0,0)[lt]{\lineheight{1.25}\smash{\begin{tabular}[t]{l}SW\end{tabular}}}}%
    \put(0,0){\includegraphics[width=\unitlength,page=2]{figures/SapphWinLabelled_Clipped.pdf}}%
  \end{picture}%
\endgroup%

         }
     \hspace{5mm}
     \subfloat[\label{fig-nozzle}]{%
         \def\svgwidth{0.24\columnwidth}
\begingroup%
  \makeatletter%
  \providecommand\color[2][]{%
    \errmessage{(Inkscape) Color is used for the text in Inkscape, but the package 'color.sty' is not loaded}%
    \renewcommand\color[2][]{}%
  }%
  \providecommand\transparent[1]{%
    \errmessage{(Inkscape) Transparency is used (non-zero) for the text in Inkscape, but the package 'transparent.sty' is not loaded}%
    \renewcommand\transparent[1]{}%
  }%
  \providecommand\rotatebox[2]{#2}%
  \newcommand*\fsize{\dimexpr\f@size pt\relax}%
  \newcommand*\lineheight[1]{\fontsize{\fsize}{#1\fsize}\selectfont}%
  \ifx\svgwidth\undefined%
    \setlength{\unitlength}{921.25984252bp}%
    \ifx\svgscale\undefined%
      \relax%
    \else%
      \setlength{\unitlength}{\unitlength * \real{\svgscale}}%
    \fi%
  \else%
    \setlength{\unitlength}{\svgwidth}%
  \fi%
  \global\let\svgwidth\undefined%
  \global\let\svgscale\undefined%
  \makeatother%
  \begin{picture}(1,1.53634306)%
    \lineheight{1}%
    \setlength\tabcolsep{0pt}%
    \put(0,0){\includegraphics[width=\unitlength,page=1]{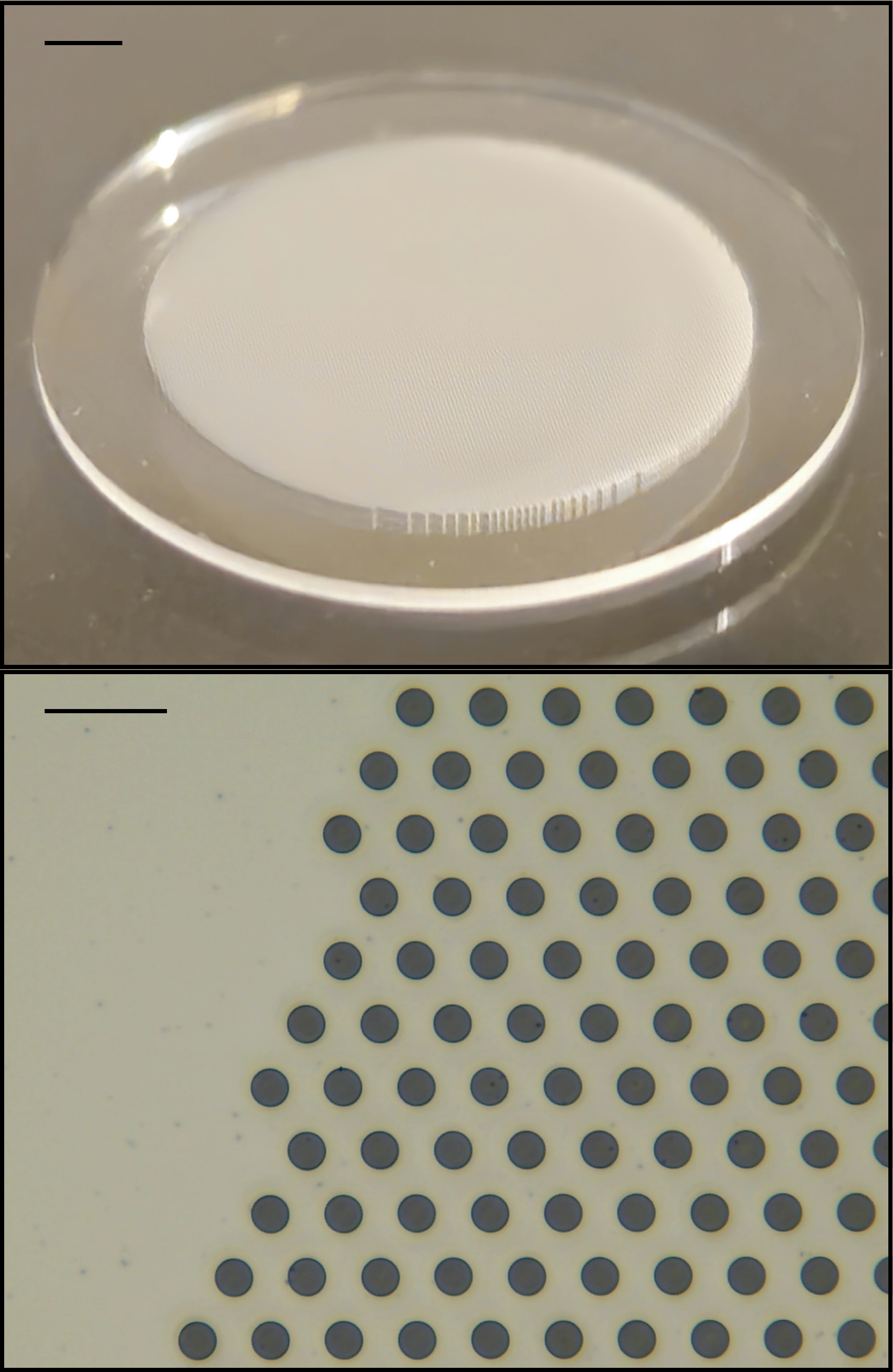}}%
    \put(0.0426885,1.417551){\color[rgb]{0,0,0}\makebox(0,0)[lt]{\lineheight{1.25}\smash{\begin{tabular}[t]{l}1mm\end{tabular}}}}%
    \put(0.05234361,0.66898166){\color[rgb]{0,0,0}\makebox(0,0)[lt]{\lineheight{1.25}\smash{\begin{tabular}[t]{l}100$\mu$m\end{tabular}}}}%
  \end{picture}%
\endgroup%

         }
     \caption{(a) Cross-section of the re-entrant oven, as oriented in the system (see Figure\ \ref{fig-fullchamber}). The top side (front) is in vacuum, with the strontium reservoir (SR) underneath the nozzle (N). The bottom side (back) is in air, with blind holes in a recess where cartridge heaters (CH) and thermocouples are inserted; the numbers mark the locations of thermocouples (see main text). (b) Cross-section of the heated sapphire window component; in this case, the bottom side, with the sapphire window (SW) itself, is in vacuum, while the top side, where the cartridge heaters (CH) are inserted from above, is in air. Grooves in the metal around the SW to allow gases to be pumped out from the inner thin-walled tube, at the top of which is a DN16CF viewport (VP) to seal the vacuum region while preserving optical access. (c) A photograph of the fused silica nozzle (top), and an optical microscope image showing the layout, size, and shape of the microchannels at the edge of the channel array (bottom).}
     \label{fig-oven-and-window}
\end{figure}

The oven (Figure\ \ref{fig-oven}) is made of three machined pieces of 316 stainless steel welded together. At the front (the top of Figure\ \ref{fig-oven}) is the oven body, with blind holes on the vacuum side where the strontium metal and nozzle sit; on the air side, heaters and thermocouples are inserted from below. At the back of the oven is a ConFlat flange that allows the oven to be attached to standard UHV systems. These pieces are connected via a thin-walled (\qty{0.25}{\milli\metre} thick) tube, designed to limit the heat conduction from the hot end of the oven to the flange, thus making the heating more efficient and keeping the exposed surfaces at safer temperatures. Numerical simulations of the thermal behaviour of the oven can be found in Appendix \ref{sec-app-comsol}.

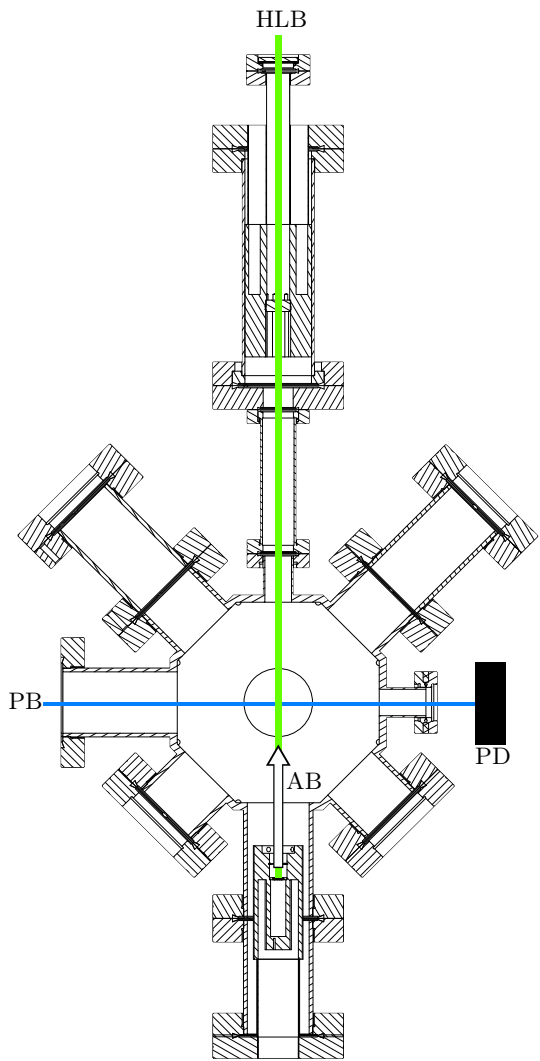
\begin{figure}[!t]
    \centering
    \def\svgwidth{0.4\textwidth}
\begingroup%
  \makeatletter%
  \providecommand\color[2][]{%
    \errmessage{(Inkscape) Color is used for the text in Inkscape, but the package 'color.sty' is not loaded}%
    \renewcommand\color[2][]{}%
  }%
  \providecommand\transparent[1]{%
    \errmessage{(Inkscape) Transparency is used (non-zero) for the text in Inkscape, but the package 'transparent.sty' is not loaded}%
    \renewcommand\transparent[1]{}%
  }%
  \providecommand\rotatebox[2]{#2}%
  \newcommand*\fsize{\dimexpr\f@size pt\relax}%
  \newcommand*\lineheight[1]{\fontsize{\fsize}{#1\fsize}\selectfont}%
  \ifx\svgwidth\undefined%
    \setlength{\unitlength}{786.84329224bp}%
    \ifx\svgscale\undefined%
      \relax%
    \else%
      \setlength{\unitlength}{\unitlength * \real{\svgscale}}%
    \fi%
  \else%
    \setlength{\unitlength}{\svgwidth}%
  \fi%
  \global\let\svgwidth\undefined%
  \global\let\svgscale\undefined%
  \makeatother%
  \begin{picture}(1,2.00468466)%
    \lineheight{1}%
    \setlength\tabcolsep{0pt}%
    \put(0,0){\includegraphics[width=\unitlength,page=1]{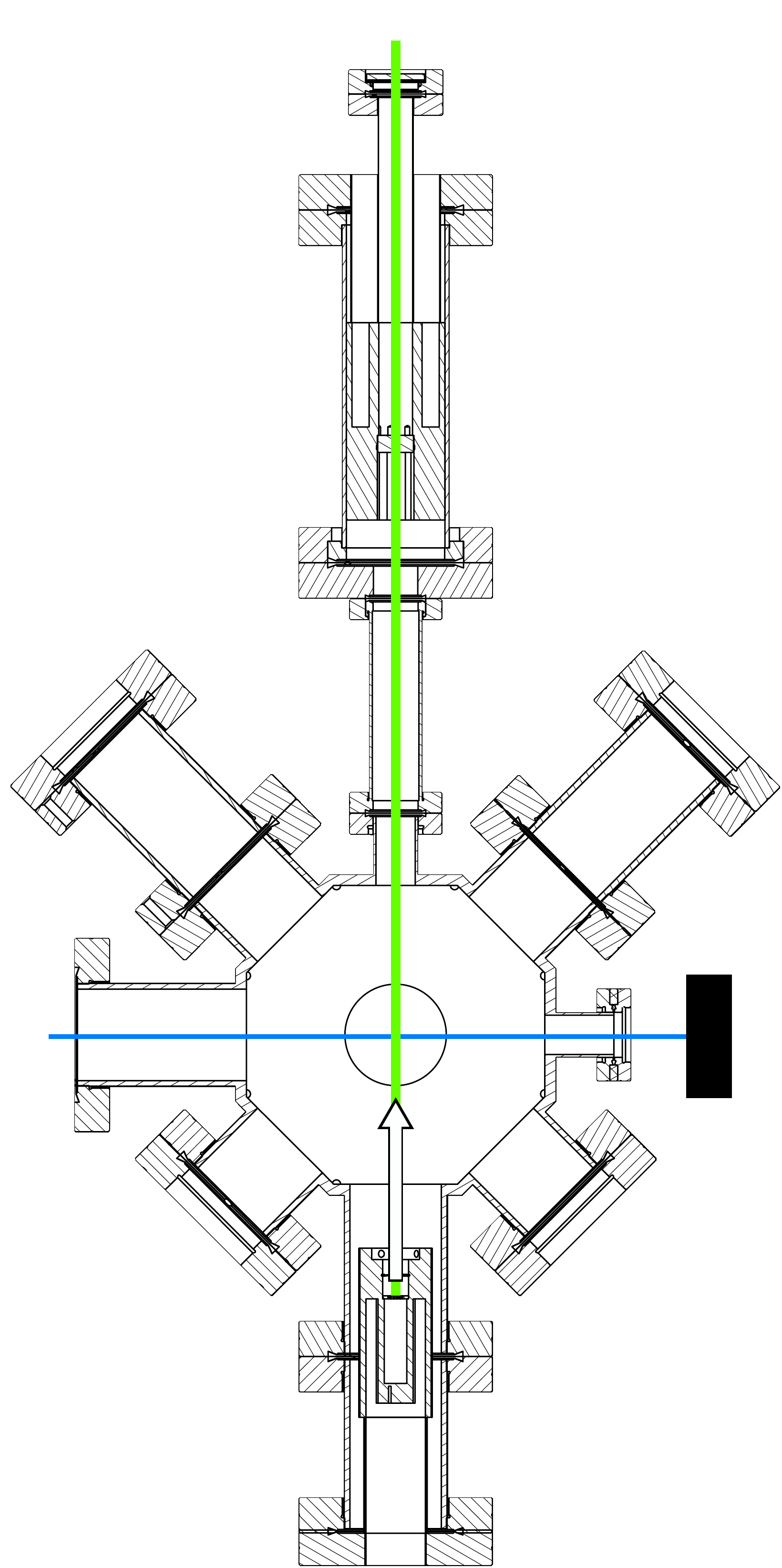}}%
    \put(-0.00392187,0.66876704){\color[rgb]{0,0,0}\makebox(0,0)[lt]{\lineheight{1.25}\smash{\begin{tabular}[t]{l}PB\end{tabular}}}}%
    \put(0.52135618,0.51739732){\color[rgb]{0,0,0}\makebox(0,0)[lt]{\lineheight{1.25}\smash{\begin{tabular}[t]{l}AB\end{tabular}}}}%
    \put(0.46373155,1.96829517){\color[rgb]{0,0,0}\makebox(0,0)[lt]{\lineheight{1.25}\smash{\begin{tabular}[t]{l}HLB\end{tabular}}}}%
    \put(0.87868757,0.56453376){\color[rgb]{0,0,0}\makebox(0,0)[lt]{\lineheight{1.25}\smash{\begin{tabular}[t]{l}PD\end{tabular}}}}%
  \end{picture}%
\endgroup%

    \caption{Cross-section of the vacuum chamber and the experimental setup, with the oven at the bottom, and the heated sapphire window at the top. The diagonal limbs provide optical access for trapping atoms in a 2D magneto-optical trap (MOT); the upper diagonal limbs are extended to ensure the viewports are not in the line-of-sight of the atomic beam. The horizontal limbs allow for a push beam, a spectroscopy probe beam, or a third cooling beam for direct loading into a 3D MOT. The heating laser beam (HLB), atomic beam (AB), probe beam (PB), and photodiode (PD) used for transverse absorption spectroscopy are labelled.}
    \label{fig-fullchamber}
\end{figure}

The re-entrant design allows the atoms to be emitted close to the centre of the chamber, and prevents the need for vacuum feed-throughs for heating that are present in traditional designs \cite{fartmann2025ramsey}. The oven body extends into the chamber from the vacuum seal, as shown in Figure\ \ref{fig-fullchamber}. The back of the oven is open to air, where six cartridge heaters (24V, 20W each) are inserted into blind holes in a regular hexagonal arrangement so that the hot ends are level with the nozzle. We attached three K-type thermocouples to the oven: (1) next to the heaters, near the nozzle; (2) at the back of the Sr reservoir; and (3) on the outside of the flange. These are marked with corresponding numbered red circles in Figure\ \ref{fig-oven}. Unless otherwise specified, references in this paper to an oven thermocouple temperature, \toven{}, are to readings from the first of these, marked 1 in Figure\ \ref{fig-oven}.

The heat from the cartridge heaters is conducted to the crucible, causing sublimation of the Sr metal therein. The design ensures the end of the crucible where the nozzle is located, closest to the hot end of the heaters, remains hotter than the Sr reservoir by an estimated 20--\qty{30}{\degree}C during normal operation. This acts to prevent clogging of the nozzle, which is an important design consideration for atomic ovens \cite{barbiero2020sideband,li2022bicolor}. Since the back of the oven is accessible during normal operation, we also have the option to further cool the crucible, e.g. with air flow, in order to increase this temperature gradient.

\begin{figure}[!t]
    \centering
    \subfloat[\label{fig-gradline}]{%
         \def\svgwidth{0.44\columnwidth}
\begingroup%
  \makeatletter%
  \providecommand\color[2][]{%
    \errmessage{(Inkscape) Color is used for the text in Inkscape, but the package 'color.sty' is not loaded}%
    \renewcommand\color[2][]{}%
  }%
  \providecommand\transparent[1]{%
    \errmessage{(Inkscape) Transparency is used (non-zero) for the text in Inkscape, but the package 'transparent.sty' is not loaded}%
    \renewcommand\transparent[1]{}%
  }%
  \providecommand\rotatebox[2]{#2}%
  \newcommand*\fsize{\dimexpr\f@size pt\relax}%
  \newcommand*\lineheight[1]{\fontsize{\fsize}{#1\fsize}\selectfont}%
  \ifx\svgwidth\undefined%
    \setlength{\unitlength}{327.4332962bp}%
    \ifx\svgscale\undefined%
      \relax%
    \else%
      \setlength{\unitlength}{\unitlength * \real{\svgscale}}%
    \fi%
  \else%
    \setlength{\unitlength}{\svgwidth}%
  \fi%
  \global\let\svgwidth\undefined%
  \global\let\svgscale\undefined%
  \makeatother%
  \begin{picture}(1,0.7648646)%
    \lineheight{1}%
    \setlength\tabcolsep{0pt}%
    \put(0,0){\includegraphics[width=\unitlength,page=1]{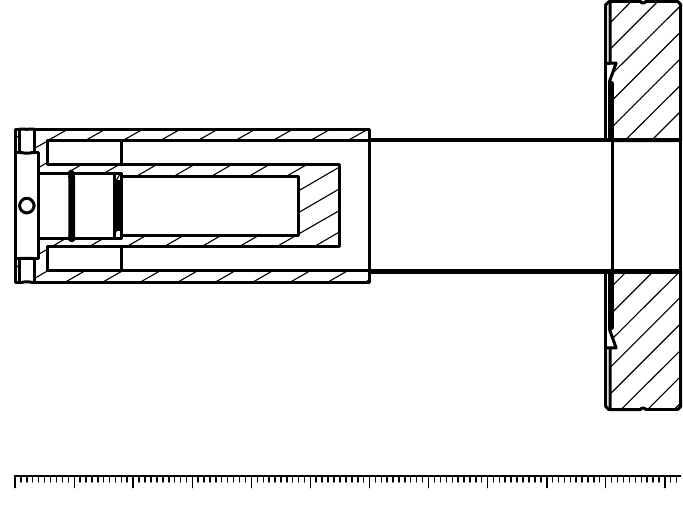}}%
    \put(-0.002043,0.00046229){\color[rgb]{0,0,0}\makebox(0,0)[lt]{\lineheight{1.25}\smash{\begin{tabular}[t]{l}0.0\end{tabular}}}}%
    \put(0.41989171,0.00046229){\color[rgb]{0,0,0}\makebox(0,0)[lt]{\lineheight{1.25}\smash{\begin{tabular}[t]{l}50.0\end{tabular}}}}%
    \put(0.84354534,0.00046229){\color[rgb]{0,0,0}\makebox(0,0)[lt]{\lineheight{1.25}\smash{\begin{tabular}[t]{l}100.0mm\end{tabular}}}}%
    \put(0,0){\includegraphics[width=\unitlength,page=2]{figures/Oven_LineDrawing_Ruled_alt.pdf}}%
  \end{picture}%
\endgroup%
%
         }
     \hfill
     \subfloat[\label{fig-tempgrad}]{%
         \includegraphics[width=0.51\columnwidth]{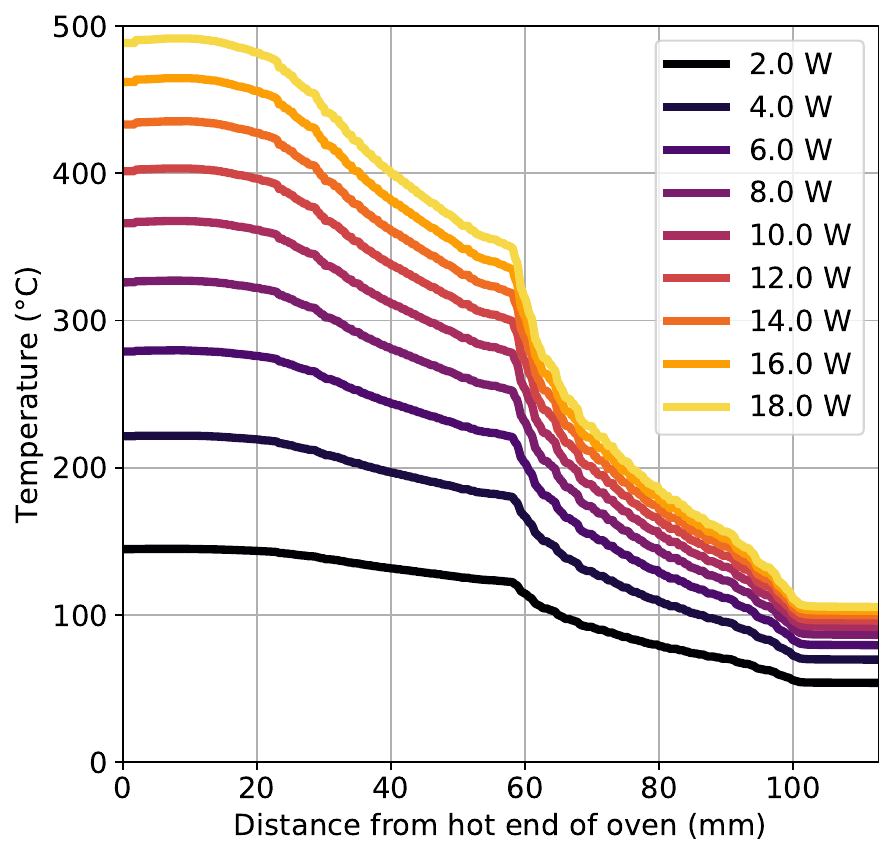}%
         }
     \hfill
    \caption{(a) Cross-section of the oven profile, with a red line highlighting the direction along which the temperature was simulated. The orange rectangles represent the hot ends of the cartridge heaters, which were the only heat sources in the model. (b) Output temperatures from the simulation as a function of position along its length as shown in (a) for a range of heater powers. This was calculated assuming only conductive and radiative losses (no convection), with a fixed effective emissivity of $\varepsilon = 0.2$ for the surfaces involved, as detailed in Appendix \ref{sec-app-comsol}.}
    \label{fig-temperature-gradient}
\end{figure}

To gain insight into the temperature distribution and the primary heat loss mechanisms in the oven, we performed thermal simulations of the oven using COMSOL Multiphysics (v5.6). The details of the modelling are explained in Appendix \ref{sec-app-comsol}. A key result of these simulations is the temperature profile along the oven body, shown in Figure\ \ref{fig-temperature-gradient}. The reduced heat conduction through the thin-walled section causes a sharp increase in the rate at which the temperature drops off at \qty{\sim60}{\milli\metre}.

Our modelling also confirmed that the heat loss from the oven is dominated by radiation at the hot (nozzle) end, and by convection at the cold (flange) end, as expected from the $T^4$ dependence of radiative heat transfer (Stefan-Boltzmann law). Comparing measured temperatures to simulations for different values of the emissivity, $\varepsilon$, for radiation, and of the heat transfer coefficient, $h$, for convection, and assuming these to be uniform throughout the oven, we estimate an effective emissivity of $\varepsilon \approx 0.2$--0.3 and a convective heat transfer coefficient of $h \approx$ 50--\qty{60}{\watt\per\metre\squared\per\kelvin} (see Appendix \ref{sec-app-comsol} and Figure\ \ref{fig-convectionradiation} therein).

Not shown in Figure\ \ref{fig-oven} nor Figure\ \ref{fig-fullchamber} -- nor explicitly included in the simulations -- is a cylindrical heat shield in place over our oven, which reduces the radiative losses. The heat shield is in vacuum, and covers the full length of the thin-walled tube and the main oven body, with a hole at the top for the atomic beam. There is also significant uncertainty in the emissivity of stainless steel (e.g. depending on the surface finish). For these reasons, we describe $\varepsilon$ as an effective emissivity; it does not represent a true emissivity for the material in the oven, but models the overall effect of these multiple factors. These simulations should therefore be considered as a qualitative guide.

Strontium atoms from the vapour inside the crucible effuse through the nozzle into the vacuum chamber. We have tested multiple different designs and materials for the nozzle, all made by machining cylindrical holes in a disc; the nozzle used in this work was fabricated using selective laser etching and micro-machining of fused silica. Previous iterations include a graphite nozzle, which collimated the atoms well, producing an angular distribution in close agreement with theoretical predictions (Appendix \ref{sec-app-lineshape-flux}) for low flux. However, the graphite nozzle could not be removed without destroying it; the fragments were observed to swell after removal, leading us to believe metal had become intercalated into the graphite. 
A stainless steel nozzle made by mechanical drilling has been shown to work well \cite{hifais} with microchannels of length $L$ = \qty{3}{\milli\metre} and diameter $d$ = \qty{0.3}{\milli\metre}, for an aspect ratio $\beta = d/L = 1/10$. This is approaching the limit of traditional methods of mechanical machining. Multichannel nozzles made from glass have been used in previous work \cite{lucas1973production,lucas2013atomic}, while glass discs with many small holes are manufactured for use in multichannel plate (MCP) detectors and have appropriate dimensions. We considered the chosen approach more controllable, as the fabrication process is customisable and silica is more robust than the glass used for MCPs.

There have been many studies of gas flow along cylindrical channels since the seminal work of Clausing \cite{clausing1971flow}, which have been used in the historical development of atomic beams \cite{ramsey1956molecular,scoles1988atomic,lucas2013atomic}. The collimation of the atom flux leaving the channels depends on this aspect ratio, $\beta$, with a smaller value of $\beta$ leading to a higher degree of collimation, but at a cost of reducing the total flux. This is because atoms that collide with the channel walls are more likely to return to the reservoir than to be passed through to the chamber \cite{senaratne2015effusive}, so the main reduction in flux is in the wings of the angular distribution; in the limit of small $\beta$, the total flux scales as $4\beta/3$ \cite{pauly2000fundamentals}. See Appendix \ref{sec-app-lineshape-flux} and \cite{pauly2000fundamentals, clausing1971flow} for a more detailed discussion of this. Other factors that inform the choice of $\beta$ include preventing clogging of the nozzle (for which a large $\beta$ is preferable) and the lifetime of the oven (for which a small $\beta$ is better), and in many cases, the convenience of fabrication.

A common way of collimating atomic beams is to stack microcapillaries to form a nozzle, for which $\beta = 1/100$ is achievable, e.g., using capillaries with $d \sim \qty{0.1}{\milli\metre}$ \cite{barbiero2020sideband}. These arrays are usually assembled by hand with generally fewer than 1000 capillaries \cite{senaratne2015effusive,barbiero2020sideband,li2022bicolor,letellier2023loading,okamoto2026direct}. This means the total flux output is limited, while clogging can become a problem. This method is also not easily scalable to large channel numbers nor for batch production.

Selective laser etching of fused silica can produce a very large number of small microchannels. The nozzle used in this work, shown in Figure\ \ref{fig-nozzle}, has \num{16213} microchannels with $d = \qty{30}{\micro\metre}$, arranged in a regular hexagonal pattern over the central \diameter\qty{8}{\milli\metre} of a disc (a filling factor of 23\%). The disc itself has \diameter\qty{10.95}{\milli\metre}, giving a border that rests on a ledge part way down the central blind hole of the oven, as shown in Figure\ \ref{fig-oven}. The disc has a thickness of \qty{300}{\micro\metre}, such that $\beta = 1/10$. The choice of parameters for this proof-of-principle testing was informed by multiple considerations but is far from the limits of what is possible with this fabrication technique. The intended use of this oven is loading of a magneto-optical trap (MOT) similar to that described in \cite{baynham2025prototype,hifais}; in our setup, this corresponds roughly to a target region of \qty{\sim1}{\centi\metre} at a distance of \qty{\sim10}{\centi\metre} from the nozzle, so $\beta = 1/10$ matches this approximately. The thickness of the silica is approaching the minimum that can easily be handled for the wafers that are etched to give batches of nozzles, and the choice of $\beta$ then determined the hole diameter for a given thickness. Other factors to consider include that a nozzle with a small thermal mass has a rapid response to heating, and very long thin channels might be more prone to clogging. Furthermore, the use of a thin disc means the channels are short, meaning the oven can be taken to higher temperatures than systems with capillaries several millimetres long before \mfp{} becomes comparable to their length, when the collimation is compromised.

\section{Heated Sapphire Window}
\label{sec-sapph}

An in-vacuum sapphire window grants optical access directly opposite the oven; the vacuum part itself is shown in Figure\ \ref{fig-sapphirewindow}, and is shown in situ at the top of Figure\ \ref{fig-fullchamber}. This allows us to shine light into the system to directly illuminate the nozzle/oven, and also provides a path for a laser beam propagating counter to the atoms, such as would be needed for a Zeeman slower.

A problem faced by windows facing high-flux atomic beams is metallization, whereby the window becomes coated in atoms. The skin depth of metals for visible light is tens of nanometres, meaning an unheated window would become opaque after a few hours of normal oven operation. We avoid metallization of the sapphire window by heating it to a temperature that ensures the rate of re-emission of atoms exceeds the rate of adsorption (cf. \cite{bowden2016adaptable,barbiero2020sideband,kobayashi2020demonstration}). This is in contrast to some other experiments, which use a large distance between window and oven \cite{li2022bicolor}, or a mirror that remains reflective when metallized \cite{huckans2018note}.

We use a design for this component that mirrors that of the oven. The $\diameter\qty{12.7}{\milli\metre}$ sapphire window sits in a position analogous to the nozzle in the oven, held in place by a circlip, and is heated using a similar arrangement of cartridge heaters. The clear aperture of the window is limited by the window itself to $\diameter$\qty{11.4}{\milli\metre} (rather than the \qty{12.0}{\milli\metre} internal diameter of the vacuum component), while the window has an anti-reflection coating on one side to optimise transmission of Zeeman slowing beams at \qty{461}{\nano\metre} such as in \cite{hifais}. The vacuum component is made from 316 stainless steel, with two thin-walled tubes and two flanges. The resulting temperature gradient ensures the sapphire window can be kept hot enough to prevent metallization while the vacuum seals remain well within their specified operating range, also reducing the risk of heat damage to the viewport both from overheating and too rapid a change in temperature (such as might occur in the event of an electrical power cut). The opposite orientation of this part means convective heat loss from the back is much more significant than for the oven, but this is not an issue, as this part does not need to reach temperatures as high as the oven.

We are able to maintain transparency of the window for oven temperatures of \toven{} $>\qty{550}{\degree}$C -- well above the normal operating temperature -- by keeping the window at $T_{\text{win}} \approx \qty{350}{\degree}$C, where $T_\text{win}$ is measured by a thermocouple in a blind hole near the sapphire window, analogous to that near the nozzle in the oven; at this temperature, the vacuum viewport on the component reaches <\qty{50}{\degree}C. While operating the oven at a fixed \toven{} = \qty{500}{\degree}C and gradually reducing $T_{\text{win}}$, we found that the window remained clear at $T_{\text{win}} = \qty{340}{\degree}$C, but began to metallize at $T_{\text{win}} = \qty{310}{\degree}$C. We deliberately allowed the window to become fully coated in Sr and opaque to visible light, then incremented $T_{\text{win}}$, demonstrating that the window became completely cleared again in \num{\sim10} hours after increasing $T_{\text{win}}$ from 310 to \qty{350}{\degree}C. This means that in the event of metallization, optical access can be restored without the need to break vacuum.

\section{Experimental Methods}
\label{sec-experiment}

We heat our oven electrically using the cartridge heaters to generate a baseline flux of Sr. High-power laser light, shone in through the heated sapphire window, heats both the nozzle and the Sr metal directly in order to quickly increase the atom flux. Because the nozzle is made of fused silica, it transmits most visible light, meaning much of the incident laser light reaches the Sr metal. This directly, locally heats the metal, increasing the sublimation rate and therefore the vapour pressure, leading to a greater flux through the nozzle. We must, however, consider whether this changes the flow regime of atoms through the nozzle, as the vapour pressure and the mean free path of the atoms, \mfp{}, will change \cite{pauly2000fundamentals} (see discussion in Section \ref{sec-results} and Appendices \ref{sec-app-lineshape-flux} \& \ref{sec-app-gaussian-mfp}).

While the nozzle is made of transparent material, the measured 77\% transmission of green light indicates significant scattering of light because of the microchannels as well as reflection, as expected surface reflectance for fused silica is $\sim$4\% per surface. The thermal mass of the nozzle is, by design, lower than that of the bulk Sr metal, meaning the nozzle is heated faster than the Sr metal, and the inherent temperature gradient is maintained. This heating of the nozzle also means that any Sr accumulated on the nozzle in steady-state will be quickly desorbed, causing an increase in the flux on a very short timescale. However, it seems likely that most of this Sr will be attached to the back of the nozzle or the walls of the microchannels, implying this only gives a small contribution to the flux which could be less well-collimated than atoms travelling through the entire length of the microchannels.

To measure the change in flux and characterise the atomic beam, we performed absorption spectroscopy. A probe laser beam transverse to the atomic beam at a distance of \qty{9.3}{\centi\metre} from the nozzle was scanned about resonance with the \qty{461}{\nano\metre} \transition{} transition in \sreights{}. The transmission of this probe beam was measured on a photodiode on the opposite side of the chamber. The setup is shown in Figure\ \ref{fig-fullchamber}. An ion pump (Agilent VacIon Plus, \qty{40}{\liter\per\second}) is connected via one of the CF40 limbs perpendicular to both laser beam paths (into the page in Figure\ \ref{fig-fullchamber}) approximately \qty{30}{\centi\metre} from the centre of the chamber. The power of the probe beam was actively stabilised to maintain a constant peak intensity of $I / I_\text{sat} = 0.20$. With this continuously running, absorption spectra were recorded and the heating laser beam was turned on and off. Some of the probe light was split off before being sent to the vacuum chamber and directed to a modulation transfer spectroscopy setup; the error signal this produced was used as a fiducial marker for the thousands of absorption features produced during each data-take.

The heating laser beam has a maximum output power of >\qty{18}{\watt} at \qty{532}{\nano\metre}, with the nominal power tunable to \qty{10}{\milli\watt} precision (produced by a Coherent Verdi V-18). We delivered the laser beam free-space, using a dichroic mirror, a low-power guide beam, and a camera to ensure the light was aligned onto the nozzle. The beam also had a diverging lens in its path so that the beam just over-filled the nozzle.

The measured power output by the heating laser at the exit aperture exceeded the nominal value set on the control unit, and we found that 75\% of the total output light reached the nozzle. For simplicity, we refer to the nominal power setting used throughout this paper. Details on the actual power reaching the nozzle and the Sr metal can be found in Appendix \ref{sec-app-power}.

With this setup, we performed experiments for heating laser powers of 5--\qty{15}{\watt}, oven thermocouple temperatures of \toven{} = 300--\qty{500}{\degree}C (near the nozzle), and heating laser pulse durations of $\Delta t = \qty{1}{\second}$ and \qty{40}{\second}. A greater heating laser power resulted in a greater increase in flux, with little evidence of saturation, so only those data taken at a power of \qty{15}{\watt} are discussed here.

\section{Results and Discussion}
\label{sec-results}

\begin{figure}[!t]
    \centering
    \includegraphics[width=\textwidth]{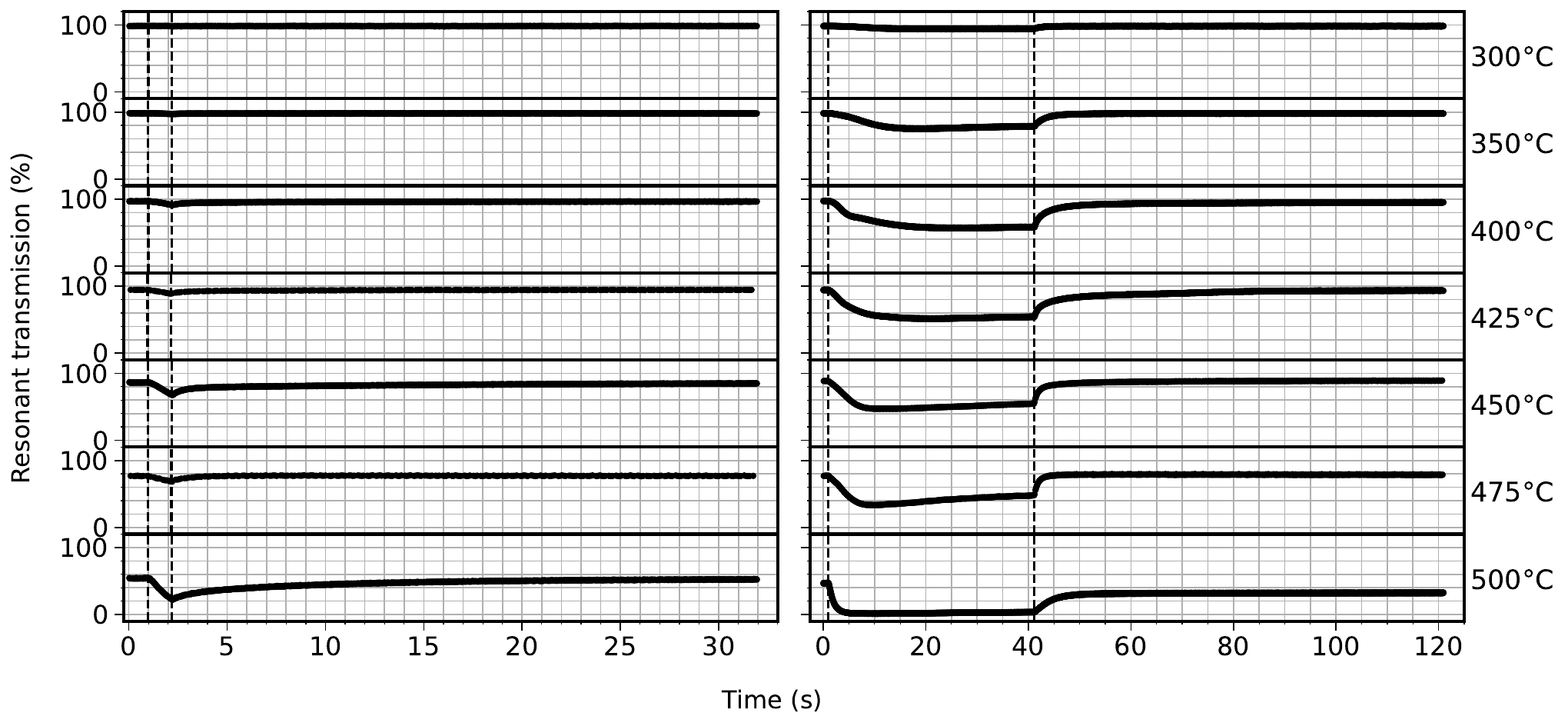}
    \caption{Resonant transmission of the probe beam as a function of time. Those plots on the left used a heating laser pulse duration of $\Delta t = \qty{1}{\second}$, while those on the right had $\Delta t = \qty{40}{\second}$. The plots are stacked vertically for different values of \toven{}, the values of which are labelled on the far right hand side. The dashed lines mark the start and end times of the heating laser pulse. In all cases shown, the heating laser light had a nominal power of \qty{15}{\watt}.}
    \label{fig-envelope}
\end{figure}

Figure\ \ref{fig-envelope} shows the resonant transmission of the probe beam through the chamber as a function of time under illumination by \qty{15}{\watt} of heating laser light for a range of values of \toven{} and for both values of $\Delta t$. The effect of the heating laser beam is to increase the flux of Sr, leading to more absorption and therefore lower transmission. The initial response to the heating light is to rapidly increase the absorption due to an increased atom flux. In certain cases, responses with two different timescales were seen -- which we attribute the fast heating of the nozzle and the slower heating of the Sr metal -- though this is not visible on the scale of Figure\ \ref{fig-envelope}; an example can be seen in Figure\ \ref{fig-envelope-example} in Appendix \ref{sec-app-envelope}, which shows the graphs for \toven{} = \qty{400}{\degree}C in more detail. In every case, the return to baseline levels is well fit by a sum of two simple exponentials, with time constants of $\mathcal{O}(1)$ \qty{}{\second} and $\mathcal{O}(10)$ \qty{}{\second}. These two different timescales could again correspond to the heating/cooling of the nozzle and the Sr metal.

For use as a method of flux modulation in experiments that create MOTs, it is important that the pressure in the chamber does not increase to prohibitively high values during and after the heating laser pulse. To this end, we monitored the pressure via the voltage output from the ion pump controller (Agilent MiniVac), and found that the most significant pressure increase when $\Delta t = \qty{1}{\second}$ came at \toven{} = \qty{500}{\degree}C, rising by a factor of $\sim$1.4 (from \qty{2e-8}{\milli\bar} to \qty{3e-8}{\milli\bar}). For all oven temperatures tested, the pressure returned to its background level -- $\mathcal{O}(\num{e-9})$ mbar at \toven{} $\leq$ \qty{475}{\degree}C -- after these short pulses within \qty{\sim1}{\second} (see Figure\ \ref{fig-pumpdata} in Appendix \ref{sec-app-pressure}).

For $\Delta t = \qty{40}{\second}$, the most significant pressure change came at \toven{} = \qty{400}{\degree}C, rising by a factor of $\sim$21 to \qty{1e-7}{\milli\bar}. For this pulse duration, the pressure returned to its background level within a few seconds of the end of the pulse, regardless of oven temperature. These results indicate that there should be no significant disruption to trap lifetimes as a result of the temporary increase in flux when using $\Delta t = \qty{1}{\second}$. There will be an effect when using longer pulses, but it quickly drops off after the end of the heating laser pulse.

\begin{figure}[!t]
    \centering
    \includegraphics[width=\textwidth]{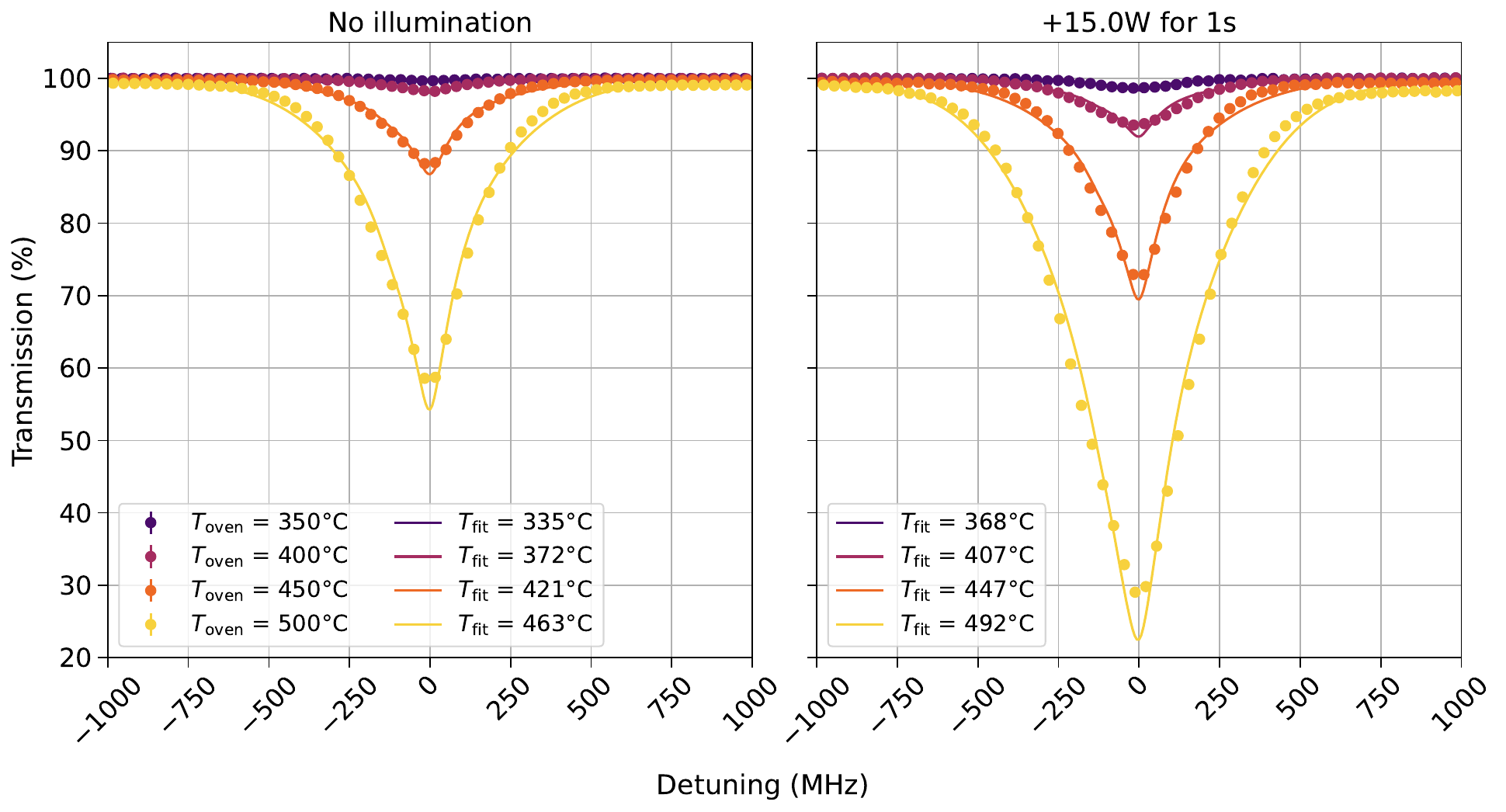}
    \caption{Transmission of the probe beam as a function of detuning from the \qty{461}{\nano\metre} \transition{} transition in \sreights{} for a range of values of \toven{}. The plot on the left shows features taken when there was no heating laser light present, while that on the right shows equivalent features taken from the end of a \qty{1}{\second} heating laser pulse. In both cases, the circular markers are experimental data points, while the solid lines are fits to the data using the theoretical lineshape for free molecular flow, given by equation \ref{eq-fit-function} in Appendix \ref{sec-app-lineshape-flux}. The heating laser light had a nominal power of \qty{15}{\watt}. The values of \toven{} in the legend on the left axes apply to the data in both plots. Also labelled are the fit vapour temperatures, $T_\mathrm{fit}$, for the theoretical lineshapes.}
    \label{fig-absorptionfeatures}
\end{figure}

In order to calculate the flux from the oven and to investigate the effect of the heating laser pulse on the collimation of the atomic beam, we look at an absorption feature from two regions of the data: one feature from the very start of each run, in the absence of any heating laser light; and one from the end of the heating laser pulse, when the system is under the maximum (for $\Delta t = \qty{1}{\second}$) or steady-state (for $\Delta t = \qty{40}{\second}$) effect of the heating laser. In every case, the absorption feature being fit has been averaged over many measurements to reduce the statistical uncertainty. The results for a selection of values of \toven{} are shown in Figure\ \ref{fig-absorptionfeatures}, where the fit vapour temperatures, $T_\mathrm{fit}$, are also labelled. We note that $T_\mathrm{fit} <$ \toven{} when the heating laser was absent, whereas $T_\mathrm{fit} >$ \toven{} after \qty{1}{\second} of illumination for oven temperatures \toven{} < \qty{450}{\degree}C.

When fitting these absorption features, we assumed the effusion to be in the free molecular flow regime (FMF, also known as the collisionless regime), which occurs when \mfp{} is much greater than the dimensions of the microchannel ($L$, $d$). In this flow regime, an analytic expression can be found for the angular distribution of atoms effusing through a cylindrical channel \cite{pauly2000fundamentals, clausing1971flow}; this distribution, $j(\theta)$, is given in Appendix \ref{sec-app-lineshape-flux}, together with details of the assumptions on which it is based. For high fluxes, we observe significant deviation from the lineshape predicted using $j(\theta)$ (discussed briefly below and in more detail in Appendix \ref{sec-app-lineshape-flux}).

The critical fit parameter is the atomic vapour temperature; using this and the geometry of the nozzle, one can calculate the total flux out of the oven. Of most interest is the flux of atoms that enter a target region of the chamber with a speed within a certain range; we refer to this as the useful flux, as these are the atoms that can be cooled and trapped for use in an experiment. The geometry of the chamber, the size of the target region, and the value of $\beta$ lead to a fixed factor of 0.388 on the flux, which is calculated as described in Appendix \ref{sec-app-lineshape-flux}. A second factor depends on two limiting values of atom speed and varies with temperature. These two factors determine the conversion from total to useful flux. We take minimum and maximum speeds of 55 and \qty{255}{\metre\per\second}, a range defined as capturable when using a Zeeman slower in a similar system \cite{hifais}. More details on the lineshape fitting and the calculation of the useful flux can be found in Appendix \ref{sec-app-lineshape-flux}.
\begin{figure}[!t]
    \centering
    \includegraphics[width=0.75\textwidth]{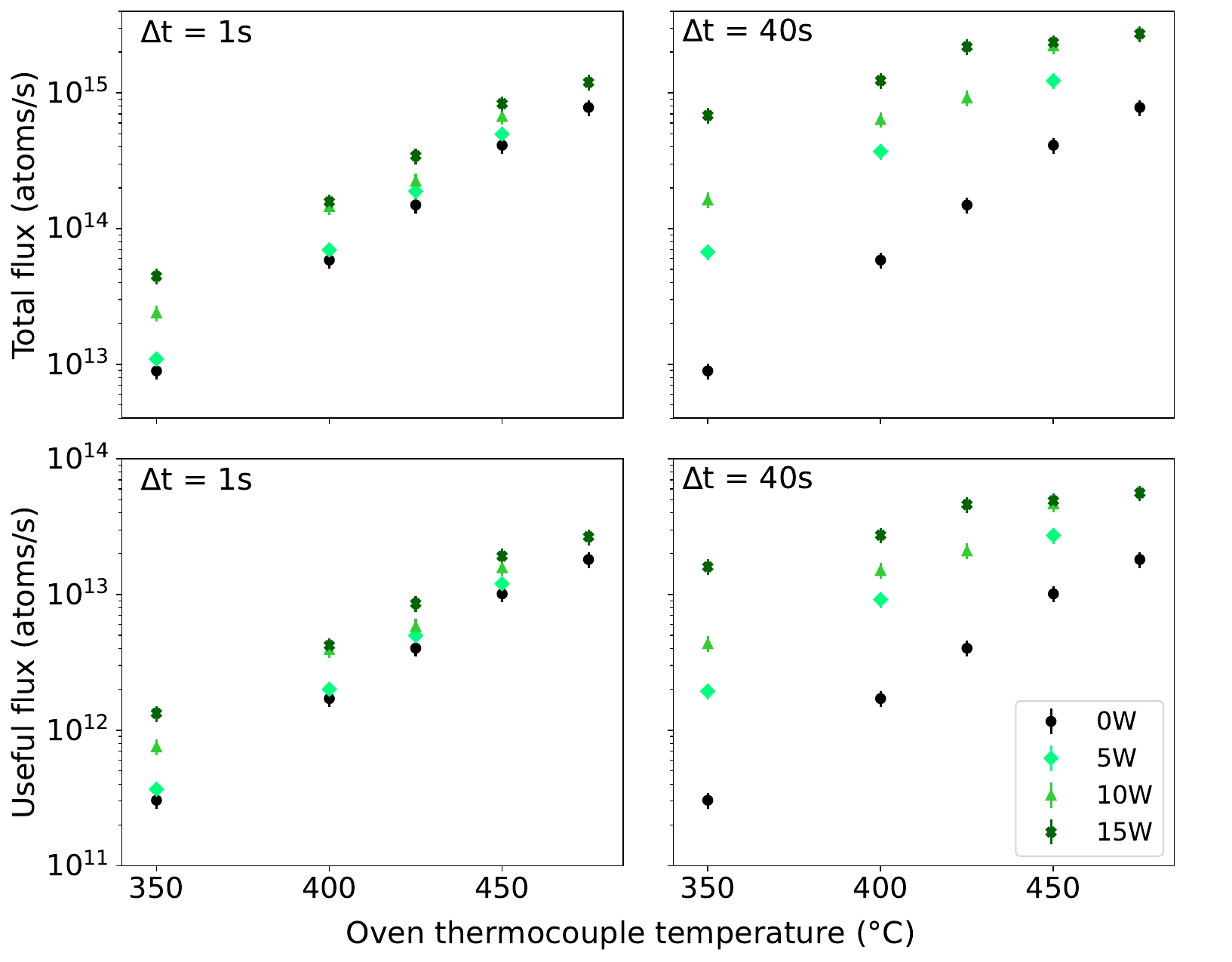}
    \caption{Flux out of the oven in atoms per second as a function of \toven{}. The graphs in the top row show the total flux out of the oven, deduced from the measured absorption and fitted lineshape, while those in the bottom row show the estimated useful flux, calculated as described in the text. The plots in the left column show the results for a heating laser pulse duration of $\Delta t = \qty{1}{\second}$, while those in the right column show the results for $\Delta t =  \qty{40}{\second}$. The legend applies to all four graphs.}
    \label{fig-fluxes}
\end{figure}

Figure\ \ref{fig-fluxes} shows the total and useful fluxes calculated from our absorption fits. We achieve a maximum total baseline flux of \num{8(1)e14} atoms/s at \toven{} = \qty{475}{\degree}C. Heating laser beam pulses of $\Delta t = \qty{1}{\second}$ and \qty{40}{\second} increase this to \num{1.2(2)e15} and \num{2.7(4)e15} atoms/s, respectively, representing increases by factors of 1.5(3) and 3.5(7). Of these, the useful flux, as defined above, increases from a baseline value of \num{1.8(2)e13} atoms/s to modulated values of \num{2.7(4)e13} and \num{5.6(7)e13} atoms/s for $\Delta t = \qty{1}{\second}$ and \qty{40}{\second}, respectively, for fractional increases of 1.5(3) and 3.1(6).
%

While these are the largest reliable flux values, the fractional effect of the heating laser beam is largest at lower values of \toven{} (and lower fluxes). At \toven{} = \qty{400}{\degree}C, the total baseline flux is \num{5.9(8)e13} atoms/s, of which \num{1.7(2)e12} atoms/s is capturable. The fractional increases in the total and useful fluxes when $\Delta t = \qty{1}{\second}$ are 2.7(5) and 2.5(5), respectively. With $\Delta t = \qty{40}{\second}$, the total and useful fluxes at \toven{} = \qty{400}{\degree}C increase by factors of 21(4) and 16(3), respectively.

The flux values calculated for \toven{} = \qty{300}{\degree}C are omitted from Figure\ \ref{fig-fluxes}, as the absorption at this temperature was negligible, meaning the results could not be meaningfully interpreted. The fluxes calculated at \toven{} = \qty{500}{\degree}C are also excluded, as the addition of the heating laser at high powers and/or long pulses resulted in a highly Gaussian lineshape, meaning the theoretical lineshape function derived assuming FMF produced a poor fit and thus an unreliable $T_\mathrm{fit}$ and flux; an example of this is shown in Appendix \ref{sec-app-gaussian-mfp}. As such, the flux values in Figure\ \ref{fig-fluxes} -- and those referenced elsewhere in the text -- are restricted to \toven{} = \qty{350}{}--\qty{475}{\degree}C. Further to this, the uncertainties on the fluxes calculated for \toven{} = \qty{350}{\degree}C are comparatively very large due to the poor signal-to-noise ratio at this temperature, so we do not discuss them beyond their inclusion in Figure\ \ref{fig-fluxes}.

The reason for the deviation of the lineshape from the theoretical FMF form towards a Gaussian at high temperatures is currently under investigation. We are running simulations to determine whether atom-atom collisions within the nozzle could result in this broadening. Collisions are believed to become significant in broadening the angular distribution without affecting the total flux if the vapour pressure is high enough that $\lambda_\text{mf} \lesssim 10L$ \cite{olander1970molecular,wouters2016design}. We are yet to confirm this as a contributing mechanism, and calculations suggest that a vapour temperature of \qty{\sim570}{\degree}C would be needed to reach this limit (though this is not a sharp transition, so even at lower temperatures and longer \mfp{}, some collisional broadening could be occurring).

We note also that, even at lower temperatures and with no heating light, the theoretical lineshape and the form of the data are consistently in slight disagreement, with some structure in the residuals (see Figure\ \ref{fig-raw-residuals} in Appendix \ref{sec-app-gaussian-mfp}). As well as the theoretical form, we also fit all data with a Gaussian, with statistically similar results, as measured with a reduced $\chi^2$ parameter (the theoretical lineshape being better only in the absence of any heating laser light -- see Figure\ \ref{fig-residuals} in Appendix \ref{sec-app-gaussian-mfp}). Despite there being no physical motivation for using a Gaussian function, it seems to work well, which suggests that our model is missing some physics, which we are currently investigating (see further discussion in Appendices \ref{sec-app-lineshape-flux} and \ref{sec-app-gaussian-mfp}).

One of the major benefits of a modulated oven flux is the increased lifetime of the source before it must be reloaded with more Sr metal. While some ultracold Sr experiments aim for continuous operation, and cycle times as low as \qty{500}{\milli\second} have been achieved \cite{bowden2019realize}, it is common for research experiments to have cycle times on the order of tens of seconds \cite{campbell2017some, thekkeppatt2025measurement}. For this reason -- and for its smaller effect on the vacuum pressure -- the results for $\Delta t = \qty{1}{\second}$ are of more significant interest than those for $\Delta t = \qty{40}{\second}$.

As an example, using this setup, the baseline flux at \toven{} = \qty{475}{\degree}C can be temporarily exceeded while operating at \toven{} = \qty{450}{\degree}C. We can perform a crude calculation of the factor by which the oven lifetime is extended by operating at this lower base temperature, where the higher flux is only produced for \qty{1}{\second} at a time. To do this, we assume the heating laser induces an instantaneous step change in the flux from the baseline level to the increased level. A cycle time of >\qty{1}{\second} is required; taking an example cycle time of \qty{10}{\second} -- and thus a duty cycle of 1/10 -- the lifetime of the oven is increased by a factor of 1.8. Even for a cycle time of \qty{1.5}{\second}, the lifetime is extended by a factor of 1.25. With oven lifetimes generally measured on a scale of years -- e.g. an oven loaded with \qty{3}{\gram} of Sr could maintain a total flux of \num{2e14} atoms/s for over three years of continuous operation -- this level of fractional change is useful and there is significant scope for improvement in future work.


\section{Conclusions}\label{sec-conc}

We have developed a compact design of a high-flux strontium oven for cold atom experiments, with a novel nozzle design and light-modulated atom flux. Selective laser etching of fused silica enables a scalable production of very small nozzles with many thousands of microchannels that produce a high degree of collimation. Verifying the robustness of the silica nozzles made by this method of manufacturing when subjected to repeated thermal cycling provides a foundation for further development. Nozzles with many more, smaller channels have been fabricated, for example with $\beta = 1/30$ ($L = \qty{300}{\micro\metre}$, $d = \qty{10}{\micro\metre}$, $N = \num{145047}$). These smaller values of $\beta$ should lead to improved collimation of the atomic beam so that a greater fraction of the flux can be cooled and trapped. With this method of fabrication, we can push to aspect ratios of $\beta \leq 1/100$ and hole diameters of $d \leq \qty{10}{\micro\metre}$.

A complementary heated sapphire window component prevents and reverses metallization of the sapphire, granting continual optical access for flux modulation and/or a Zeeman slowing beam, also allowing visual inspection of the nozzle. We demonstrated reversal of metallization of the window without breaking vacuum.\footnote{
\begin{minipage}[t]{\linewidth}
\normalsize{Enquiries about the supply of custom vacuum components can be directed to \href{mailto:christopher.foot@physics.ox.ac.uk}{christopher.foot@physics.ox.ac.uk}.}
\end{minipage}
}

We modulated the atom flux by direct illumination of the nozzle and oven with high-power laser light, and found that the system responded to this rapidly. A pulse duration of $\Delta t = \qty{1}{\second}$ and a heating laser power of \qty{15}{\watt} increased the useful flux by a factor of 2.5(5) at an oven thermocouple temperature of \toven{} = \qty{400}{\degree}C, falling to a factor of 1.5(3) at \qty{475}{\degree}C. We found that with $\Delta t = \qty{40}{\second}$, these factors increased to 16(3) at \qty{400}{\degree}C and 3.1(6) at \qty{475}{\degree}C. This effect allows the oven to be run at a lower base temperature while temporarily achieving a higher flux, extending the lifetime of the oven.

We have observed reasonable agreement with theoretical expectations of absorptive lineshape in an effusive atomic beam up to a point. We found that using $\Delta t = \qty{40}{\second}$ at \toven{} = \qty{500}{\degree}C caused the absorption lineshape to deviate from the lineshape expected in the free molecular flow regime, appearing to take on a more Gaussian profile. While we have hypotheses for the mechanism behind this, it remains an open issue that we are actively investigating.

Other future work includes looking into alternative heating light sources. Diode-pumped solid state lasers are not an appropriate first choice for this application, being bulky, energy-intensive, and excessively expensive. Furthermore, heating does not require single-mode laser light. One alternative we are considering is to use a multimode laser diode array, such as the Nichia NUBM3NT, which can output up to \qty{200}{\watt} centred on \qty{455}{\nano\metre}. This is a more economically viable laser, although it might not be possible to collect all of the output and direct it onto the nozzle. Besides the lower cost and easier installation/operation, this increase in power should yield a faster response. This would enable this approach to be used in experiments with shorter cycle times and would reduce the required base temperature, further increasing the oven lifetime. The use of blue heating light at a wavelength similar to the \qty{461}{\nano\metre} light required for laser cooling of strontium is convenient with regards the optics and anti-reflection coatings that would be used for a Zeeman slower. High-power lasers are available at other wavelengths, so the dependence of the flux modulation effect on the wavelength of the heating laser light could also be investigated.


\section{Acknowledgements}\label{sec-ack}

We thank Dr. Dharmalingam Prabhakaran for help loading the oven and installing the nozzle. We thank Dr. Richard Hobson and Prof. John Ellis for useful discussions. We acknowledge use of COMSOL Multiphysics through a concurrent licence in Oxford Physics. This work was supported by the EPSRC IAA’s STFC highlight fund, grant reference EP/X525777/1; by UKRI through its Quantum Technology for Fundamental Physics programme, via the Grant No. ST/T006633/1 from STFC in the framework of the AION Consortium; and by EPSRC grant reference EP/Y004175/1. J.S. acknowledges support from the Rhodes Trust. 


\newpage

\onecolumngrid

\appendix
\label{sec-appendix}

\section{Thermal simulations of the oven}
\label{sec-app-comsol}

\begin{figure}[!t]
     \centering
     \subfloat[\label{fig-radiationboundaries}]{%
         \includegraphics[width=0.18\textwidth]{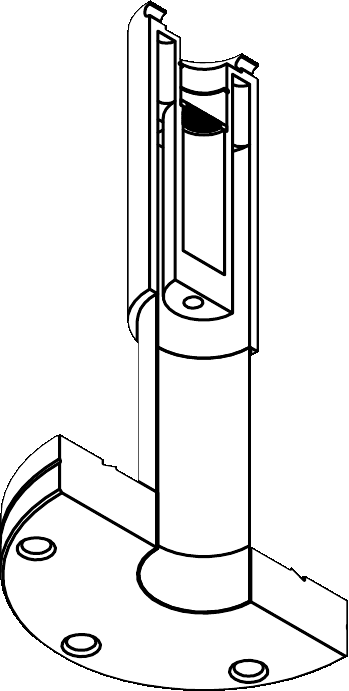}%
         }
     \hspace{10mm}
     \subfloat[\label{fig-convectionboundaries}]{%
         \includegraphics[width=0.18\textwidth]{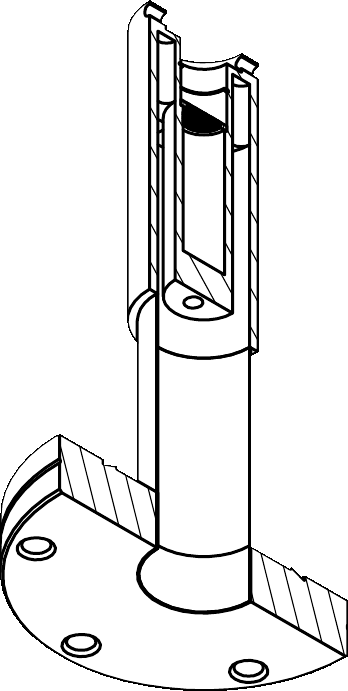}%
         }
     \caption{CAD model of the oven, where the green highlighted surfaces are those with non-zero thermal loss mechanisms in the COMSOL simulations. (a) The outward-facing boundaries, which were assigned non-zero emissivity, $\varepsilon$, when considering radiative losses. (b) Those surfaces in open air, which were assigned non-zero heat transfer coefficient, $h$, for convective losses. In both cases, all relevant surfaces are visible from this viewing angle (e.g. the top side of the flange always had $\varepsilon = h = 0$).}
     \label{fig-comsolboundaries}
\end{figure}

We used finite element analysis to carry out thermal simulations of the oven in COMSOL. In these simulations, heat conduction was always present within single solid bodies (all three constituent parts of the oven are together considered a single solid), while conduction between bodies was only enabled between the heaters and the surrounding oven body. The hot ends of the heaters were modelled as six cylinders of length \qty{12.5}{\milli\metre} and diameter \qty{4.1}{\milli\metre}; these served as the only source of heat in the simulations. Radiative and convective heat loss processes between boundaries were switched on and off to investigate their influence; Figure\ \ref{fig-comsolboundaries} shows the surfaces that were used for each mechanism. We compared the results of these simulations to measured temperatures taken using thermocouples attached to the oven (see Figure\ \ref{fig-oven}).

Modelling radiative processes rigorously would require ray tracing of multiple reflections, so we made the simplifying assumption that all outward facing surfaces of the oven radiate away energy into a room-temperature background, and varied the surface emissivity, $\varepsilon$, which we treated as being the same for all surfaces highlighted in Figure\ \ref{fig-radiationboundaries}. This is also done to account for the uncertainty in $\varepsilon$ arising from the fact that the emissivity of stainless steel depends on several factors, such as whether the surface is polished or tarnished. The internal surfaces of the oven are assumed to have no radiative loss ($\varepsilon = 0$), as we do not expect the general thermal behaviour of the oven to depend strongly on these surfaces, which have smaller surface areas than the outer ones and which are largely surrounded by surfaces of a similar temperature (as opposed to the room temperature surroundings of the external faces). Note, however, that the outer surface of the crucible containing the Sr metal (see Figures\ \ref{fig-oven} \& \ref{fig-radiationboundaries}) is assigned $\varepsilon > 0$, despite being largely surrounded by the stainless steel oven body. This part of the oven is designed to have a temperature gradient, which can only be achieved in steady state when some loss mechanism is present.

Convection loss inside the open-to-air blind hole where the heaters/thermocouples are inserted is considered to be zero, since the orientation of the oven means hot air will remain trapped in that space. We thus take heat loss by convection to occur only via the outer surfaces of the ConFlat flange (back and side boundaries, as shown in Figure\ \ref{fig-convectionboundaries}), as these are the only surfaces not in vacuum around which there is free circulation of air.

\begin{figure}[!t]
     \centering
     \subfloat[\label{fig-convectionflange}]{%
         \includegraphics[width=0.48\columnwidth]{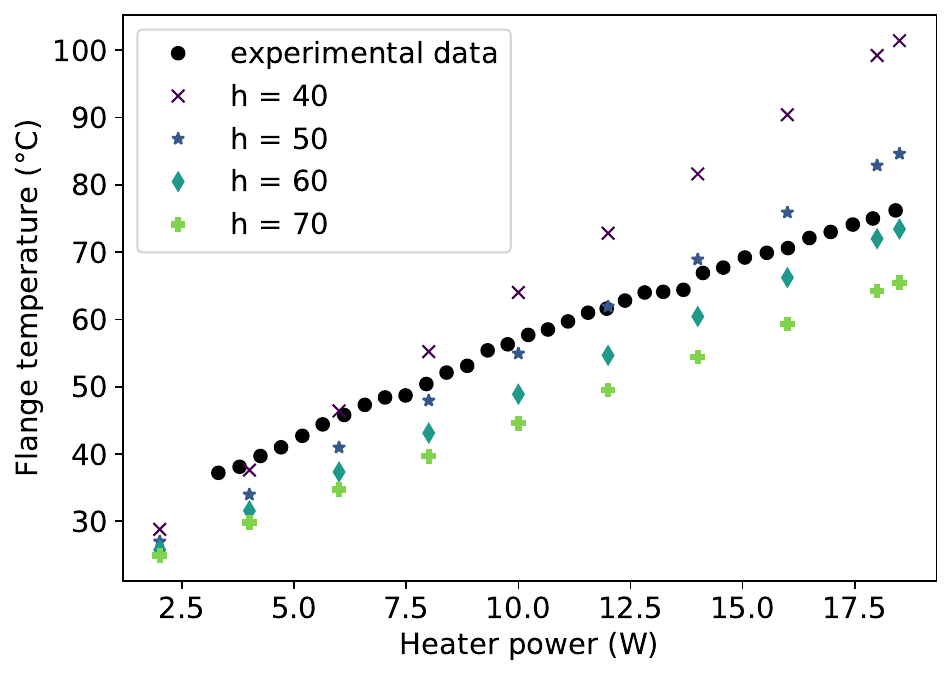}%
         }\hfill
     \subfloat[\label{fig-radiationnozzle}]{%
         \includegraphics[width=0.48\columnwidth]{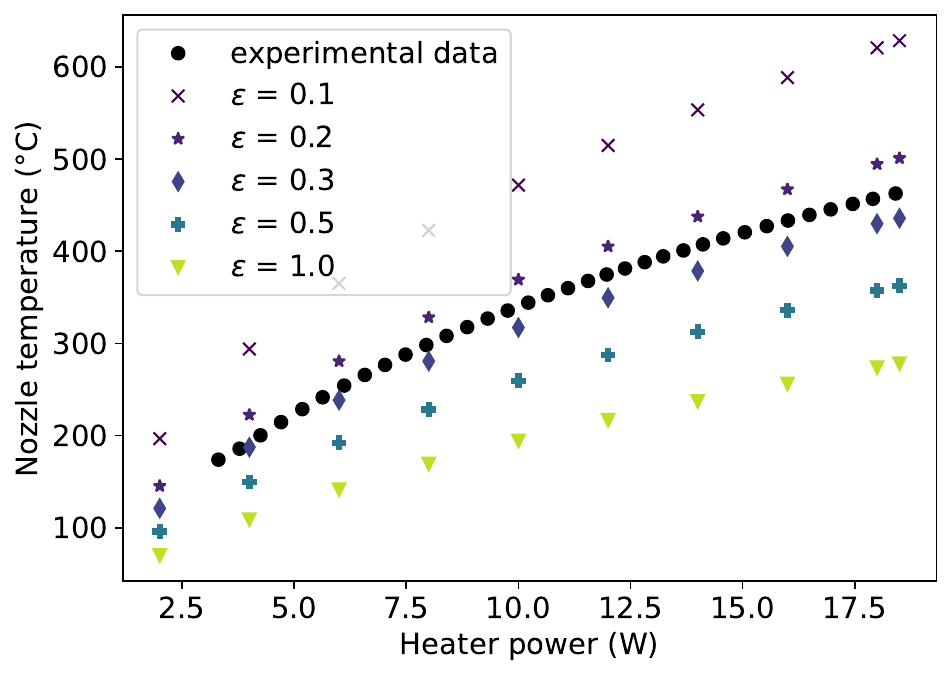}%
         }
     \caption{Comparison of simulated and measured temperatures of the flange and the nozzle for various values of the heat transfer coefficient, $h$, and the emissivity, $\varepsilon$. (a) Measured flange temperature alongside simulated flange temperature when only conductive and convective losses were present (no radiation) for a range of $h$ values. (b) Measured nozzle temperature alongside simulated nozzle temperatures with only conductive and radiative losses (no convection) for a range of $\varepsilon$ values. The values of $h$ in the legend in (a) are in units of \qty{}{\watt\per\metre\squared\per\kelvin}, while $\varepsilon$ in (b) is dimensionless.}
     \label{fig-convectionradiation}
\end{figure}

We varied the simulated power delivered by the heaters and extracted the simulated temperature of various parts of the oven. Firstly, we did this for purely conductive and convective losses (no radiation) for a range of values for the convective heat transfer coefficient, $h$, which is assumed to be the same for all surfaces highlighted in Figure\ \ref{fig-convectionboundaries}. The results for the flange temperature are shown in Figure\ \ref{fig-convectionflange}. The measured flange temperature exhibits a similar trend and slope to the simulated results for $h =$ 50--\qty{60}{\watt\per\metre\squared\per\kelvin}. Under these conditions, the simulated nozzle temperature ranged from 450\qty{\sim3000}{\degree}C, clearly not agreeing with the measured temperatures shown in Figure\ \ref{fig-radiationnozzle}.

Indeed, we next considered only conductive and radiative modes of heat loss (no convection), and ran the simulation for different values of $\varepsilon$. Figure\ \ref{fig-radiationnozzle} shows the results, comparing the simulated to the experimentally measured nozzle temperature. In this case, the simulated results for $\varepsilon = $ 0.2--0.3 are in close agreement with the measured nozzle temperature, confirming that radiation is the dominant mechanism of heat loss at the hot end of the oven. This is as expected, since the Stefan-Boltzmann law for radiated power depends on temperature as $T^4$, meaning it should become more significant than convection at higher temperatures.

With radiation enabled and convection suppressed, the simulated flange temperatures have values quite similar to the measured values shown in Figure\ \ref{fig-convectionflange}, however the largely linear dependence of the measured flange temperature on heater power did not match the sublinear trend of the radiative simulations. We therefore conclude that the flange temperature is primarily governed by convective losses.
%

During our COMSOL simulations, another area we investigated was the temperature gradient along the oven as a whole. While the oven is designed to be hotter at the nozzle than at the back of the Sr reservoir, it is also designed to have a gradient along its full length in order to maximize its efficiency, keeping the flange cool to minimize losses from this end. Figure\ \ref{fig-gradline} in the main text shows a cross-section of the oven with the line along which the simulated temperature was sampled shown in red; meanwhile, Figure\ \ref{fig-tempgrad} shows this simulated surface temperature as a function of position along this line for a range of heater powers. The results shown in Figure\ \ref{fig-temperature-gradient} assumed a fixed effective emissivity of $\varepsilon = 0.2$ for all radiating surfaces, and with no convective losses. We found that even at oven thermocouple temperatures of $\sim$\qty{500}{\degree}C, the flange remains at \qty{\sim100}{\degree}C. This is beneficial for energy efficiency, ensures compatibility with systems in which the oven would be close to components that are sensitive to high temperatures/temperature gradients, and is safer for users, as the flange is easily accessible.

\section{The heating laser power reaching the nozzle/oven}
\label{sec-app-power}

In the main text, the quoted power of the heating laser beam is the nominal power set on the control unit of the laser. The measured power output at the exit aperture is in excess of this, and there are losses as the beam passes through various optics in the beam path. There is further power lost due to clipping after expanding the beam. The main text refers to the nominal value for simplicity, but here we provide details of the measured output power and calculations of the power that reaches the nozzle and that reaches the Sr metal.

Table \ref{tab-powers} shows the nominal powers, the measured output powers, and the powers reaching/passing through the nozzle. The loss upon reflection off the mirrors and transmission through the diverging lens in the beam path was found to be negligible, as those components all had appropriate dielectric coatings. Losses through three objects therefore contribute to the decrease in power along the beam path: the dichroic mirror used to direct the beam into the chamber while allowing us to image the nozzle; the vacuum viewport; and the sapphire window. These are measured to transmit 89\%, 92\%, and 92\% of \qty{532}{\nano\metre} light, respectively, with no significant variation at different laser powers.
\begin{table}
    \centering
    \begin{tabular}{|c|c|c|c|}
        \hline
        Nominal Power (W) & Actual Power at Aperture (W) & Power at Nozzle (W) & Power at Sr Metal (W) \\ \hline
        15.00 & 16.60 & 10.88 & 8.38 \\
        12.50 & 13.72 & 8.99 & 6.92 \\
        10.00 & 10.91 & 7.15 & 5.51 \\
        7.50 & 8.14 & 5.33 & 4.11 \\
        5.00 & 5.37 & 3.52 & 2.71 \\
        2.50 & 2.66 & 1.74 & 1.34 \\
        1.00 & 1.03 & 0.68 & 0.52 \\ \hline
    \end{tabular}
    \caption{Nominal heating laser powers, as quoted in the main text, compared to the measured output power of the laser and the power that reached the nozzle and the Sr metal. This takes into account the transmission loss due to optical elements (and nozzle) in the beam path and the clipping due to the expansion of the beam.}
    \label{tab-powers}
\end{table}

The values for the power at the nozzle are scaled by a further 87\% due to the power lost by clipping of the expanded beam. When installing the diverging lens, we measured 87\% of the power after the lens to be within the \qty{10.95}{\milli\metre} diameter of the nozzle at the same distance as the nozzle would be from the lens. The power reaching the Sr metal is then 77\% of this, which was the measured transmission of the fused silica nozzle at \qty{532}{\nano\metre}.

Experiments were performed using a range of heating laser powers, generally for nominal values of 5.00, 10.00, and \qty{15.00}{\watt}, with a small number at the other powers listed in Table \ref{tab-powers}. The main text discusses results primarily for a nominal \qty{15}{\watt} (though with data taken at 5 and \qty{10}{\watt} included in Figure\ \ref{fig-fluxes}). This is because the trends with heating laser power were as expected, in that a greater power resulted in a greater increase in flux, so only those data taken at the highest nominal power are reported. The results in the main text therefore pertain to a power of \qty{10.88}{\watt} incident on the nozzle, of which \qty{8.38}{\watt} is transmitted to the Sr metal.

\section{Response time of the system to the heating laser light}
\label{sec-app-envelope}

\begin{figure}[t!]
    \centering
    \includegraphics[width=0.7\linewidth]{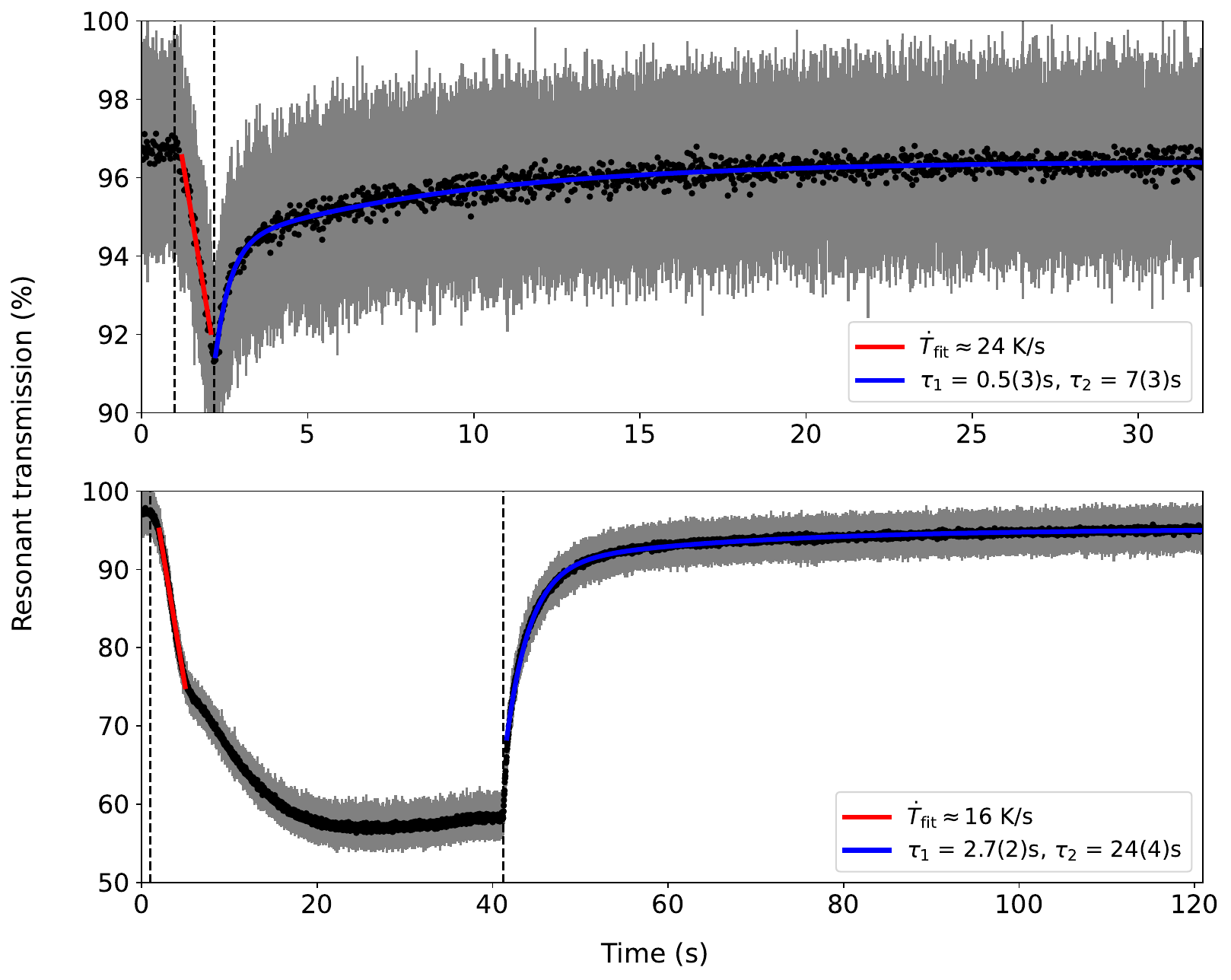}
    \caption{Resonant transmission of the absorption spectra at \toven{} = \qty{400}{\degree}C for $\Delta t = \qty{1}{\second}$ (top) and $\Delta t = \qty{40}{\second}$ (bottom). In both cases, the initial response to the introduction of the heating laser light has been fit with a simple linear function, shown in red, which has been matched to a rate of increase in the fit vapour temperature, $\dot{T}_\mathrm{fit}$. The response to the removal of the heating laser light has been fit with the sum of two simple exponentials with time constants $\tau_1$ and $\tau_2$, shown in blue. The black points are the average of the measured data, while the grey shows the standard deviation. The dashed lines mark the start and end of the heating laser pulse, which had a nominal power of \qty{15}{\watt}.}
    \label{fig-envelope-example}
\end{figure}

In Figure\ \ref{fig-envelope} in the main text, we showed the resonant transmission of the probe beam for a range of oven thermocouple temperatures, \toven{}, and for two heating laser pulse durations, $\Delta t$. While this illustrates the overall trend in the response of the system to the heating laser light, details are not easily seen on the scale of that figure, for example the timescale of the response. Figure\ \ref{fig-envelope-example} shows the resonant transmission for the specific case of \toven{} = \qty{400}{\degree}C. These are the same data that are shown for this temperature in Figure\ \ref{fig-envelope}.

Also included in Figure\ \ref{fig-envelope-example} are fits to two regions of the data. The initial response is fit with a linear function of the form $mx + c$, shown in red in Figure\ \ref{fig-envelope-example}. This linear fit has then been converted to a rate of increase in the fit vapour temperature, $\dot{T}_\mathrm{fit}$, in kelvin per second. The calculated rates give insight into the timescale on which the system responds to the heating laser light, and demonstrates that this method is considerably faster than modulating the electrical heater power. The value of $\dot{T}_\mathrm{fit}$ is greater for the smaller $\Delta t$, as it is fit to a region of lower absorption and thus lower fit temperature, where we expect the rate of heating to be faster.

The response to the removal of the heating light is fit for both $\Delta t$ with a superposition of two simple exponentials, ($A + Be^{-(t-t_0)/\tau_1} + Ce^{-(t-t_0)/\tau_2}$), shown in blue in Figure\ \ref{fig-envelope-example}. This is to account for the thermal relaxation of the nozzle and the Sr metal at different rates. The two times constants are separated by approximately an order of magnitude, with the longer timescale of this response on the order of $\mathcal{O}(10)$ \qty{}{\second}. This further confirms that this method of flux modulation is significantly faster than changing the electrical heater power to achieve the same flux increase, which happens on a scale of $\mathcal{O}(1)$ \qty{}{\hour}.

Note that for the case of $\Delta t = \qty{40}{\second}$ (the lower plot in in Figure\ \ref{fig-envelope-example}), there are clearly at least two timescales to the response to the introduction of the heating light, with a fast initial response followed by a slower levelling off. We attribute this to the fast heating of the nozzle and the slower heating of the Sr metal.

\section{Changes in chamber pressure due to the heating laser}
\label{sec-app-pressure}

\begin{figure}[t!]
    \includegraphics[width=\columnwidth]{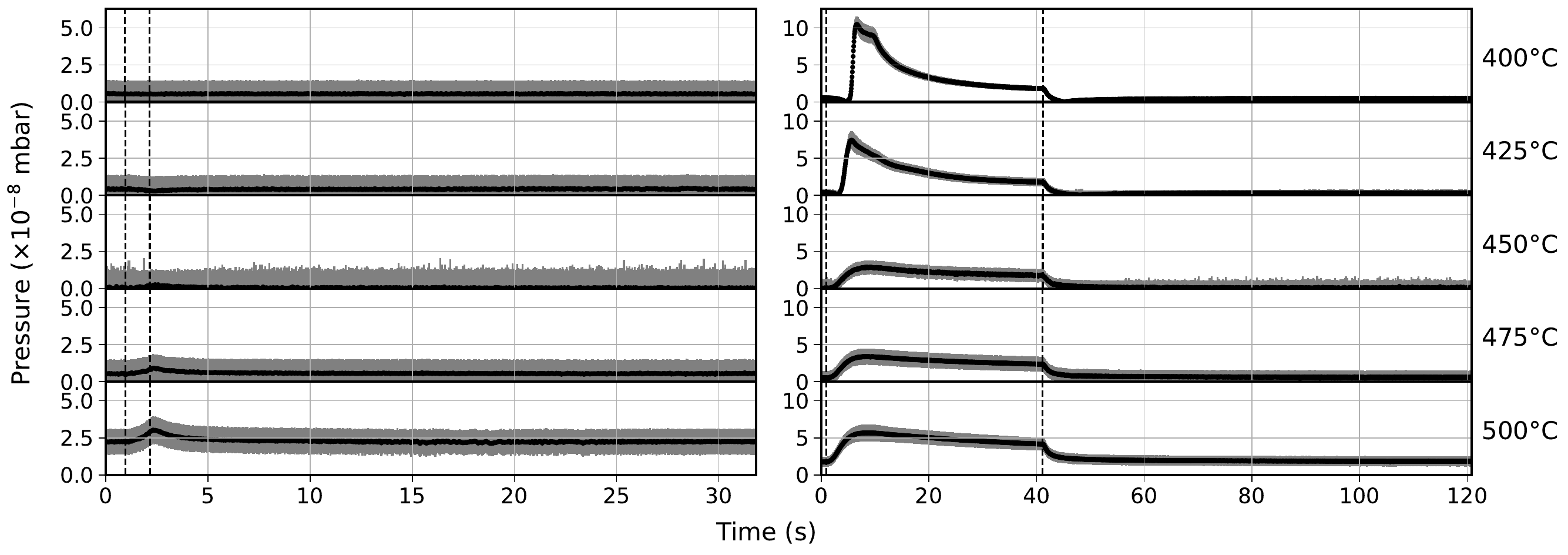}
    \caption{The pressure in the vacuum chamber plotted as a function of time. The graphs on the left hand side show the results for a heating laser pulse duration of $\Delta t = \qty{1}{\second}$; those on the right hand side show the results for $\Delta t = \qty{40}{\second}$. The plots are vertically stacked depending on oven thermocouple temperature, \toven{}, the values of which are shown on the far right hand side. The solid black lines mark the average values, while the grey regions mark the $1\sigma$ uncertainties. The dashed lines mark the start and end of the heating laser pulses.}
\label{fig-pumpdata}
\end{figure}

The vacuum in the chamber was maintained using an Agilent VacIon Plus 40 Diode ion pump (\qty{40}{\liter\per\second}), controlled with an Agilent MiniVac. The output signal from this pump controller was sampled simultaneously with the photodiodes used in the spectroscopy in order to monitor the effect of the heating laser beam on the chamber pressure. The results are shown in Figure\ \ref{fig-pumpdata}. The background pressure was $\lesssim\qty{5e-9}{\milli\bar}$ for \toven{} $\leq$ \qty{475}{\degree}C, increasing considerably to \qty{\sim2e-8}{\milli\bar} at \toven{} = \qty{500}{\degree}C. We note that these values do not represent the exact pressure at the centre of the chamber, however they give insight into the changes in that pressure due to the heating laser light. With a \qty{21}{\centi\metre} CF40 limb connecting the pump to the chamber, we estimate the conductance to the pump to be \qty{20}{\liter\per\second}, meaning the pumping rate of the ion pump is not the limiting factor.

The effect of the heating laser when operated with $\Delta t = \qty{1}{\second}$ is very small, with the largest change in chamber pressure coming at \toven{} = \qty{500}{\degree}C, where the pressure rose from \qty{2e-8}{\milli\bar} to \qty{3e-8}{\milli\bar}, increasing by a factor of $\sim$1.4. With $\Delta t = \qty{40}{\second}$, the largest effect came at \toven{} = \qty{400}{\degree}C. In this case, the pressure rose from \qty{5e-9}{\milli\bar} to \qty{1e-7}{\milli\bar}, an increase by a factor of $\sim$21. Nevertheless, the data for the two pulse durations are in agreement, as, while the response is faster at higher temperatures, the pressure reaches greater maximum values at lower temperatures if given the time.

Note in particular that, with $\Delta t = \qty{1}{\second}$, the pressure returns to its background level on a timescale of $\qty{\sim1}{\second}$, even for the largest pressure increases. This means that trap lifetimes are unlikely to be significantly affected by these heating laser pulses, which is another advantage over modulating via increased electrical heater power. Even with $\Delta t = \qty{40}{\second}$, this recovery is on a similar timescale of $\mathcal{O}(1)$ \qty{}{\second}, so trap lifetimes are likely to be affected only for the duration of the pulse.

\section{Lineshape fitting and flux calculations}
\label{sec-app-lineshape-flux}

The absorption lineshape is fit using the Beer-Lambert law for the intensity of light transmitted through an absorptive medium, which is a simple exponential relation, $I = I_0e^{-D}$, for optical depth $D$. This optical depth is calculated using the kinetic theory of particles effusing through a cylindrical channel in combination with the scattering cross-section. The value of $D$ for \sreights{} is given by,
\begin{equation}
    D_{88} = \frac{\eta}{l}\int_{-\theta_\text{lim}}^{\theta_\text{lim}}d\theta\int_0^\infty dv \frac{f_b(v,T)j(\theta)\sigma_0}{v\left( 1 + \frac{I}{I_{\text{sat}}} + \frac{4(\delta_{88}-kv\sin(\theta))^2}{\Gamma^2} \right)},
\end{equation}
where $\eta$ is the peak intensity of the atomic beam (in atoms per second per steradian); $l = \qty{9.3}{\centi\metre}$ is the shortest distance from the nozzle to the probe beam; $\theta_\text{lim} = \qty{477}{\milli\radian}$ is the maximum half-angle subtended by an atom emitted from the oven that will interact with the probe beam (as determined by the chamber geometry and measured from the surface normal to the nozzle); $f_b(v, T)$ is the distribution of speeds, $v$, of particles in a beam of temperature $T$; $j(\theta)$ is the angular distribution of atoms emitted from the nozzle; $\sigma_0 = 3\lambda_0^2/2\pi$ is the resonant cross-section for a transition of wavelength $\lambda_0$ \cite{footbook} (and taking $1/\lambda_0 = \qty{21698.452}{\per\centi\metre}$ \cite{NIST-ASD}); $I$ is the intensity of the probe beam (in \qty{}{\watt\per\metre\squared}); $I_\text{sat} = \qty{42.7}{\milli\watt\per\centi\metre\squared}$ is the saturation intensity of the transition; $\delta_{88}$ is the angular detuning of the probe beam from resonance with the transition in \sreights{}; $k = 2\pi/\lambda$ is the wavenumber of the probe beam; and  $\Gamma = 2\pi\times\qty{32}{\mega\hertz}$ is the natural linewidth of the transition \cite{NIST-ASD}.

The first of these values, $\eta$, is found as the product of the total open area of the nozzle and the maximum flux density through a single channel (equal to the flux density entering a channel, $\phi_\text{in}$), divided by the total solid angle of a hemisphere (as only atoms transmitted out one side of the nozzle can interact with the probe beam), and multiplied by 2 (analogous to the peak intensity of a Gaussian):
\begin{equation} \label{eq-eta}
    \eta = Nr^2\phi_\text{in} = Nr^2\frac{p(T)}{\sqrt{2\pi mk_BT}},
\end{equation}
where $r = d/2$ is the microchannel radius and $p(T)$ is the vapour pressure of strontium, for which we use the empirical formula found in \cite{pucher2025srreferencedata}. Then $f_b(v, T)$ is the modified Maxwell-Boltzmann speed distribution for particles in a beam \cite{footbook},
\begin{equation}
    f_b(v, T) = \frac{2v^3}{v_p^4}e^{-v^2/v_p^2},
\end{equation}
where $v_p = \sqrt{2k_BT/m}$ is the most probable speed of a particle of mass $m$ in a gas of temperature $T$. The angular distribution, $j(\theta)$, of atoms effusing through a cylindrical channel in the free molecular flow (FMF) regime is given by \cite{pauly2000fundamentals},
\begin{equation} \label{eq-jtheta}
    j(\theta) = \alpha\cos(\theta) + \frac{2}{\pi}\cos(\theta) \left( (1-\alpha)\left(\arccos(q_{\text{min}}) - q\sqrt{(1-q_{\text{min}}^2)}\right) + \frac{2}{3q}(1-2\alpha)\left(1 - (1-q_{\text{min}}^2)^{3/2}\right) \right),
\end{equation}
with,
\begin{align} \label{eq-alpha}
    \alpha &= \frac{1}{2} - \frac{1}{3\beta^2}\cdot\frac{1 - 2\beta^3 + (2\beta^2 - 1)\sqrt{1 + \beta^2}}{\sqrt{1 + \beta^2} - \beta^2\text{arsinh}\left(\frac{1}{\beta}\right)}\\
    q &= \frac{\tan{\theta}}{\beta}\\
    q_{\text{min}} &= \text{min}(1, q).
\end{align}
As in the main text, $\beta = d/L$ is the aspect ratio of the diameter to the length of the channel. This formulation separates the distribution by angle: $q < 1$ for small angles subtended by the atoms and $q > 1$ for large angles. Alternative formulations separate the distribution into atoms that pass straight through the channel and those that collide with the wall at least once \cite{tschersich1998formation}.

We fit the data using a superposition of six of these lineshapes, one for each hyperfine component of each stable isotope of strontium. These are weighted according to the relative abundances of the isotopes, $A_n$ \cite{nist2010srdata}, and with the relative strengths of the different hyperfine contributions from the fermionic $^{87}$Sr isotope, $S_{(F)}$ \cite{ye2013production}:
\begin{equation}\label{eq-fit-function}
    \frac{I}{I_0} = A_{88}e^{-D_{88}} + A_{86}e^{-D_{86}} + A_{84}e^{-D_{84}} + A_{87}\left(S_{(7/2)}e^{-D_{87}^{(7/2)}} + S_{(9/2)}e^{-D_{87}^{(9/2)}} + S_{(11/2)}e^{-D_{87}^{(11/2)}}\right).
\end{equation}
When calculating the optical depths, $D_n$ (or $D_{87}^{(F)}$ for $^{87}$Sr), the laser detuning values, $\delta_n$ ($\delta_{87}^{(F)}$), are offset by the corresponding isotope shift \cite{ye2013production}, acting to shift the centres of the individual lineshapes. The vapour temperature, $T$, is the only physical fit parameter, which determines the width and depth of the lineshape. Other fit parameters are simply linear and constant offset terms which we use to fix the far off-resonance baseline to a uniform 100\% transmission.

This fit function relies on several assumptions \cite{clausing1971flow,pauly2000fundamentals,olander1970molecular,wouters2016design}. All atoms colliding with the channel walls are assumed to undergo fully diffuse reflections, meaning all atom-wall collisions result in the atom being re-emitted according to Lambert's cosine law (although partially specular reflections are considered likely to occur \cite{clausing1971flow}, and could affect the distribution of emitted atoms); the density of atoms is assumed to decrease linearly along the length of the channels; the effusion is treated as being in the collisionless regime of free molecular flow such that atom-atom collisions are not considered. This means that the velocity distribution is separable into uncoupled speed and angular distributions. The function also models all atoms as being emitted from the centre of the microchannel array, simply scaling the result for one channel by the number of channels. The finite extent of the array will affect the velocity distribution of atoms interacting with the probe beam (e.g. those atoms passing straight through a channel with $\theta = 0$ would interact with the probe beam only for those channels on the line directly underneath it); we expect this effect to be small due to the relative scales of the microchannel array ($\diameter\qty{8}{\milli\metre}$) and the distance to the probe beam ($\geq\qty{9.3}{\centi\metre}$). These simplifications might be contributing to the imperfect fit to the data.

Once the lineshape has been fit and a vapour temperature extracted, the total flux coming out of the oven can be calculated as,
\begin{equation}
    \Phi_\text{out} = N\pi r^2W\phi_\text{in},
\end{equation}
where $N\pi r^2$ is the total open area of the nozzle, while $W$ is the probability that a particle entering a cylinder of aspect ratio $\beta$ is transmitted out the other end, and $\phi_\text{in}$ is the flux density entering a single microchannel as given in equation \ref{eq-eta} above. $W$, also known as the Clausing factor, is given by \cite{pauly2000fundamentals},
\begin{equation}
    W = 1 + \frac{2}{3}(1 - 2\alpha)\left(\beta-\sqrt{1+\beta^2}\right) + \frac{2}{3\beta^2}(1+\alpha)\left(1-\sqrt{1+\beta^2}\right),
\end{equation}
where $\alpha$ is defined in equation \ref{eq-alpha}. This allows us to calculate the total flux in atoms per second that exits the nozzle into the chamber. 

While the total flux itself is of some interest, the more pertinent figure of merit is the flux that can feasibly be cooled and trapped, since not all of the flux is useable in laser cooling applications \cite{baynham2025prototype,hifais}. MOT sizes depend on the fraction of atoms that enter a target region of the chamber and that have speed below a certain capture velocity so that they can be sufficiently cooled to be trapped. Slow atoms are more sensitive to collisions than faster ones, implying there is a maximum number density in the strontium reservoir -- corresponding to a limit on the vapour pressure and temperature -- that can be used to form a useful effusive atomic beam. In the FMF regime, and in the limit of $\beta<<1$, $W$ tends to $4\beta/3$, i.e. the total flux of atoms from a cylindrical channel is a factor of $4\beta/3$ times the rate of effusion from a hole of the same diameter (a channel of zero length) \cite{pauly2000fundamentals}. We have $4\beta/3 = 0.133$ corresponding to a reduction of the total flux by a factor of $7.5$ for the same forward intensity of atoms. This arises because atoms that collide with the walls are in fact more likely to return to the reservoir rather than be transmitted through to the chamber \cite{senaratne2015effusive}.
\begin{figure}[t!]
    \centering
    \includegraphics[width=0.5\linewidth]{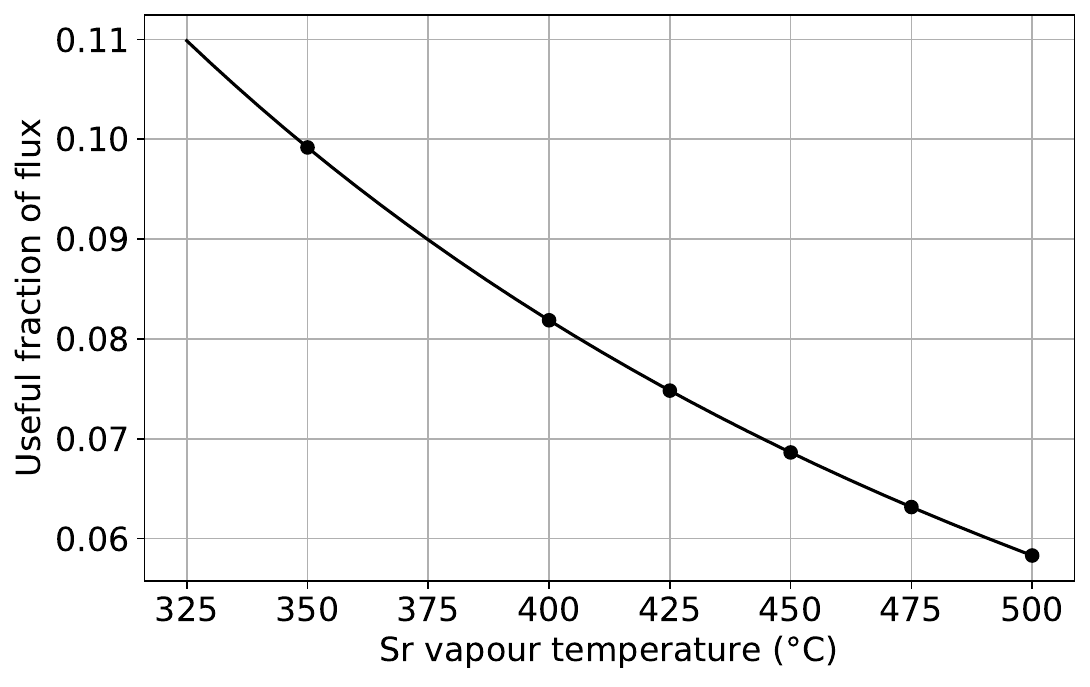}
    \caption{The fraction of the flux that could feasibly be captured in a MOT as a function of the vapour temperature. This fraction is calculated as the product of $f_1$ and $f_2$, as defined in the text; this temperature-dependence is only due to $f_2$. The line shows the functional form of this fraction, while the circles mark the values corresponding to the calculated fluxes in Figure\ \ref{fig-fluxes}.}
    \label{fig-frac-vs-temp}
\end{figure}

The useful fraction of the flux is calculated by scaling the total output flux with two factors. The first of these, $f_1$, corresponds to the fraction of atoms emitted that will enter the capture region of the chamber. This factor depends only on the geometry of the system, and is the same in all cases for a given experimental setup. For the chamber used in this work, we assume a cooling beam diameter of \qty{25}{\milli\metre} (as this will not be clipped by standard DN40CF viewports). The geometry of our chamber then means the maximum half-angle subtended by an atom emitted from the nozzle that will reach the intersection of these beams is \qty{0.192}{\radian}. The value of $f_1$ is then,
\begin{equation}
    f_1 = \frac{\int_0^{0.192}d\theta \sin{(\theta)}j(\theta)}{\int_0^{\pi/2}d\theta \sin{(\theta)}j(\theta)},
\end{equation}
which implicitly depends on the aspect ratio of the nozzle microchannels, $\beta$, through $j(\theta)$.

The second factor, $f_2$, corresponds to the fraction of atoms that have speeds within some range that could feasibly be captured by laser cooling. This factor depends on the atom vapour temperature, and therefore on an extracted fit parameter. With the atoms taken to have the speed distribution $f_b(v,T)$ given above, $f_2$ is calculated as:
\begin{equation}
    f_2 = \frac{\int_{v_{\text{min}}}^{v_{\text{max}}}dv f_b(v, T)}{\int_0^\infty dv f_b(v, T)},
\end{equation}
where $v_\text{min}$ and $v_\text{max}$ are, respectively, the minimum and maximum speeds of an atom that could be captured. In our flux calculations, we take $v_\text{min} = \qty{55}{\metre\per\second}$ and $v_\text{max} = \qty{255}{\metre\per\second}$ based on results from a similar system with a Zeeman slower \cite{hifais}.

The useful flux is finally calculated as $\Phi_\text{useful} = f_1f_2\Phi_\text{out}$. Figure\ \ref{fig-frac-vs-temp} shows this scaling factor, $f_1f_2$ as a function of the Sr vapour temperature, highlighting the decreased efficiency at higher temperatures due to more of the atoms exceeding the maximum capture speed.

\section{Deviation from the theoretical lineshape}
\label{sec-app-gaussian-mfp}

\begin{figure}
    \centering
    \includegraphics[width=0.75\linewidth]{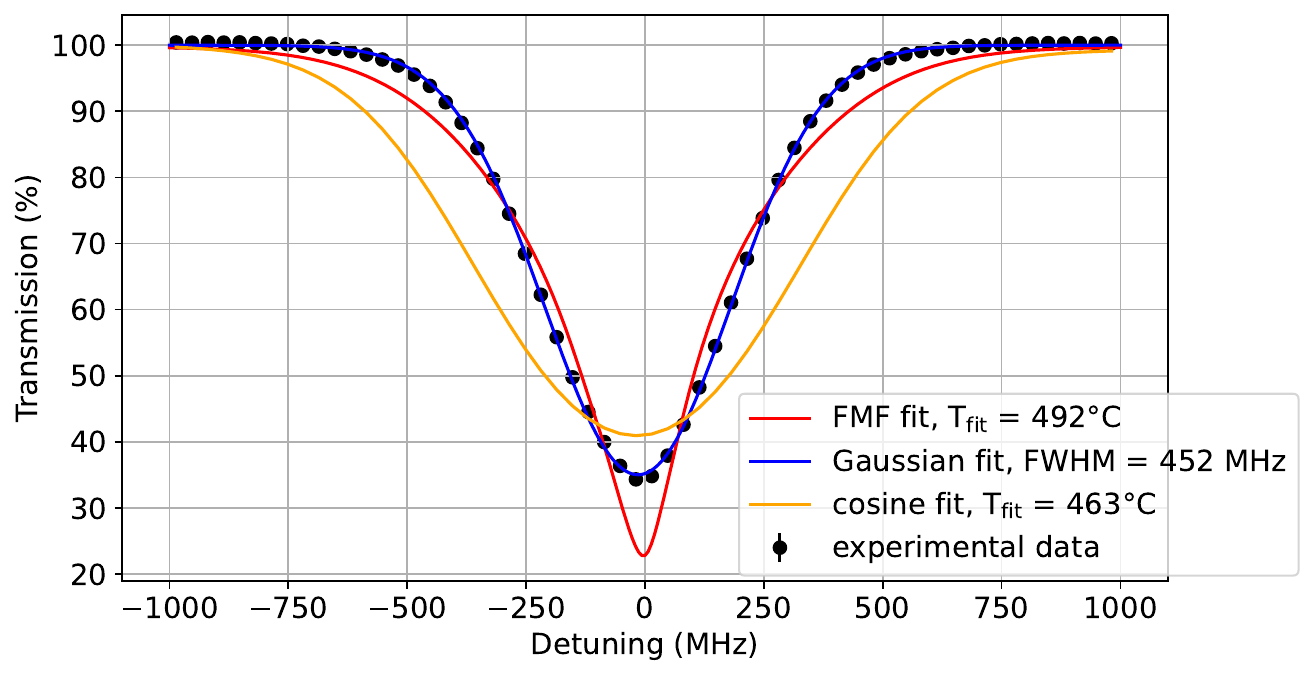}
    \caption{Absorption data taken at \toven{} = \qty{500}{\degree}C and at the end of a \qty{40}{\second} pulse of \qty{15}{\watt} heating laser light. Shown in red is a fit to the data using the theoretical effusive lineshape for free molecular flow (FMF) using $j(\theta)$ as defined in equation \ref{eq-jtheta}; in blue is a fit using a simple Gaussian function; and in orange is a theoretical fit assuming no collimation (e.g. due to entirely specular atom-wall reflections), replacing $j(\theta)$ with $\cos(\theta)$.}
    \label{fig-gaussian}
\end{figure}

As \toven{}, $\Delta t$, and the power of the heating laser beam are increased, the lineshape of the transverse absorption profile begins to deviate from the theoretical lineshape for FMF, becoming distinctly more Gaussian. An illustrative example of this is seen in Figure\ \ref{fig-gaussian}, showing the data taken at \toven{} = \qty{500}{\degree}C and $\Delta t = \qty{40}{\second}$, with a nominal heating laser beam power of \qty{15}{\watt}. Qualitatively, the Gaussian appears to fit the data better than the theoretical effusive lineshape. In order to quantitatively confirm this and investigate the transition between flow regimes, we look at the reduced $\chi^2$ parameters for the two fit functions.

\begin{figure}[!t]
     \centering
     \subfloat[\label{fig-residuals}]{%
        \includegraphics[width=0.95\linewidth]{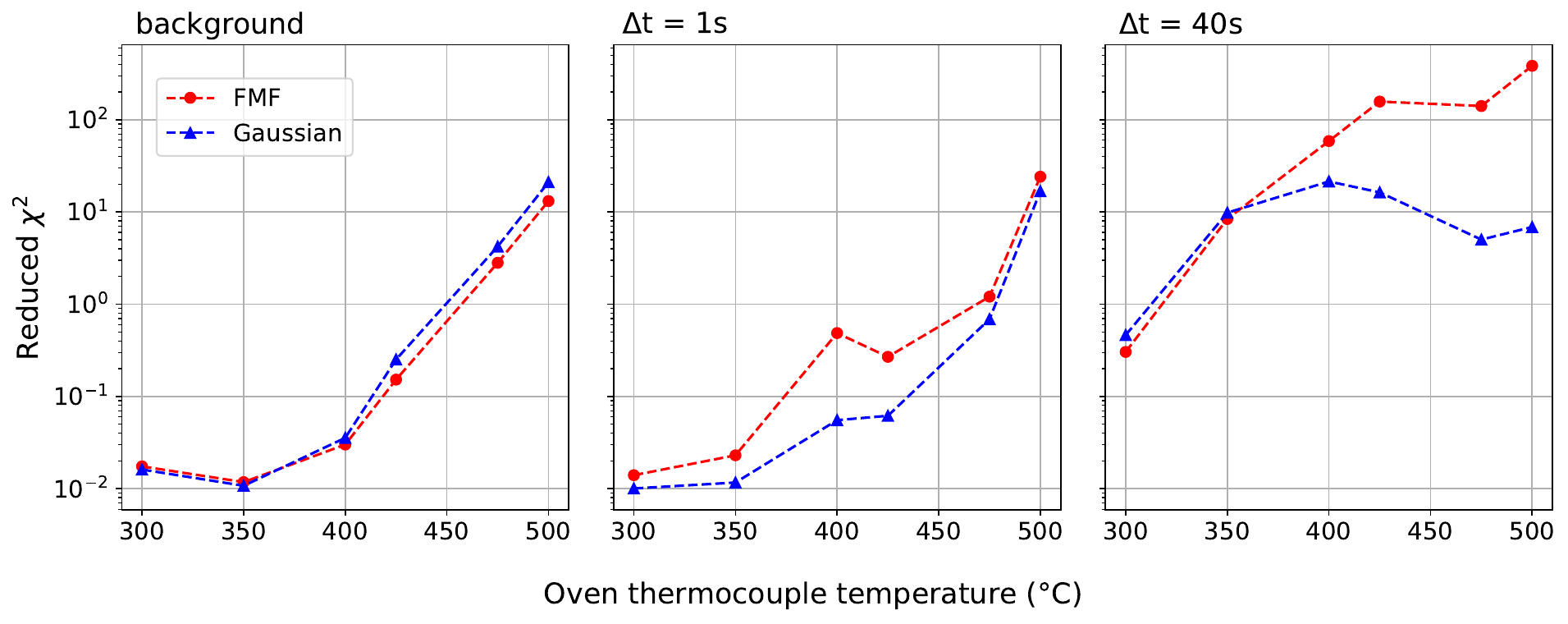}
        }
     \\
     \subfloat[\label{fig-raw-residuals}]{%
        \includegraphics[width=0.95\linewidth]{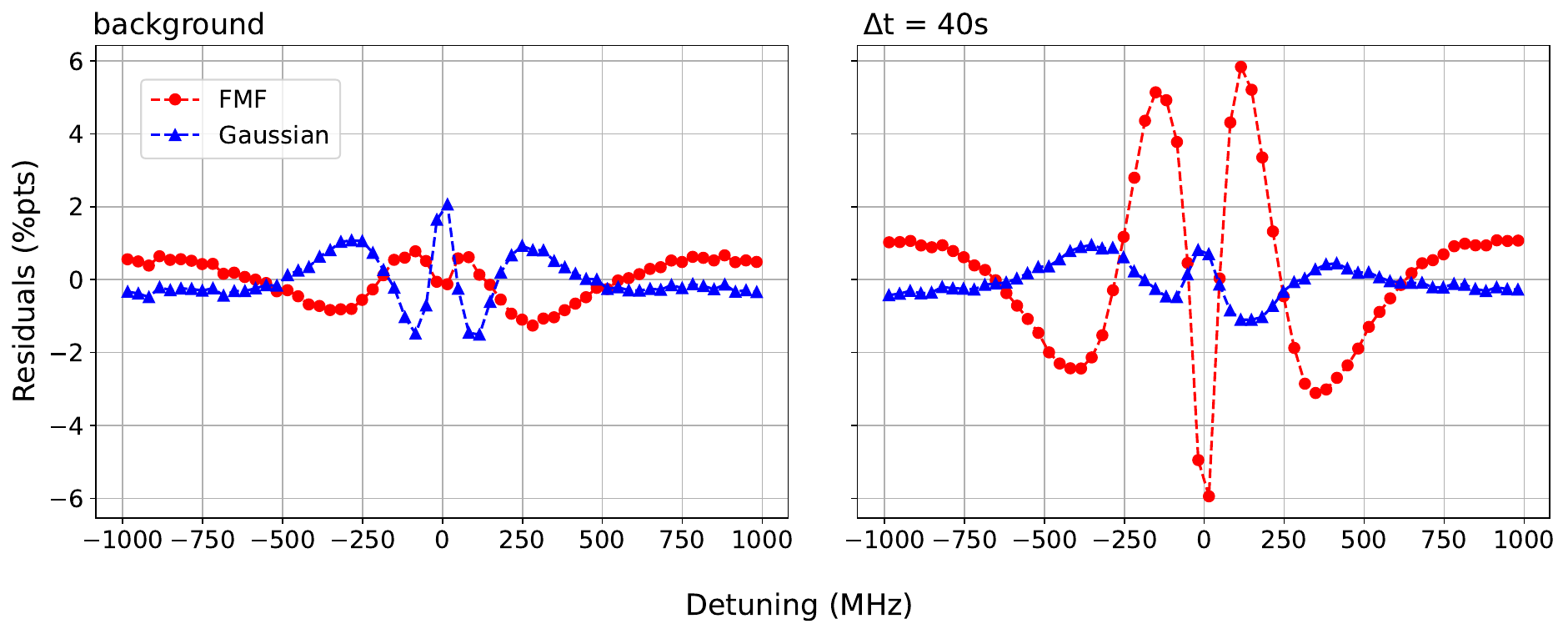}
        }
     \caption{(a) Reduced $\chi^2$ parameters as a function of \toven{}, showing that the Gaussian fit is statistically better than the theoretical lineshape for free molecular flow (FMF) at high \toven{} and long heating laser pulses. (b) Raw residuals for the two fit functions at \toven{} = \qty{475}{\degree}C; the left plot shows the residuals for the fits to the baseline feature, with no heating laser light, while that on the right shows those for the fits to the feature taken after \qty{40}{\second} of illumination. In all graphs, the dashed lines are a guide for the eye.}
     \label{fig-bothresiduals}
\end{figure}

Figure\ \ref{fig-residuals} shows the reduced $\chi^2$ parameter as a function of \toven{} in three regimes: baseline absorption, with no heating laser light; at the end of a \qty{1}{\second} pulse; and at the end of a \qty{40}{\second} pulse. It is clear from Figure\ \ref{fig-residuals} that, by this metric, a Gaussian fit slightly worse than the theoretical lineshape fit to the baseline absorption features, but slightly better after \qty{1}{\second} pulses of heating laser light; in both these cases, the trend of the reduced $\chi^2$ with \toven{} is the same for both fit functions. It is with $\Delta t = \qty{40}{\second}$ that the Gaussian clearly becomes a better fit at high temperatures, at least according to this measure. In this region, the reduced $\chi^2$ decreases as \toven{} increases, which is the opposite trend to all other data in Figure\ \ref{fig-residuals}.
%
%
%

This is concerning given that there is no physical motivation to fit the data with a Gaussian function. It suggests that the long-pulse modulation of the flux, while effective at increasing the total flux, compromises the collimating power of the nozzle. This means the fraction of that flux which could be trapped and used in experiments is reduced. Due in part to this (and in part to the need for faster cycle times), the main text focusses more on the results from those experiments that used $\Delta t = \qty{1}{\second}$.

Figure\ \ref{fig-raw-residuals} then shows the raw residuals for \toven{} = \qty{475}{\degree}C for the baseline feature and after $\Delta t = \qty{40}{\second}$. This confirms what is seen in Figure\ \ref{fig-residuals}, where the theoretical FMF fit is marginally better than the Gaussian fit when the heating laser light is absent, but becomes significantly worse than the Gaussian after \qty{40}{\second} of illumination. Figure\ \ref{fig-raw-residuals} shows us that this is a result of both the FMF model becoming worse and the Gaussian becoming better. The graphs also show some structure in the residuals, with the two fit functions having the same structure but with opposing signs. This adds to our belief that the model used is incomplete.

One avenue that we are currently investigating as a possible cause of this transition is the presence of atom-atom collisions within the nozzle. As \mfp{} decreases towards the length scales of the nozzle, a transition from FMF to other flow regimes occurs. If \mfp{} is less than ten times the microchannel length ($10L$ = \qty{3}{\milli\metre} in this work), then the flow enters the opaque regime wherein interatomic collisions between atoms travelling axially along the channel become significant, which is expected to broaden the angular distribution without reducing the total flux \cite{olander1970molecular,wouters2016design}. In the extreme case where \mfp{} is less than the diameter of the nozzle microchannels ($d$ = \qty{30}{\micro\metre} in this work), then the flow becomes viscous rather than kinetic as interatomic collisions occur between atoms travelling across the channel \cite{olander1970molecular,wouters2016design,pauly2000fundamentals}. This leads to hydrodynamic effects and is expected to reduce the total flux as well as affecting the angular distribution.

With \mfp{} = $k_BT/\left(\sqrt{2}\pi d_0^2 p(T)\right)$ for an atomic vapour at a temperature $T$ and a pressure $p(T)$, and atoms having an effective diameter $d_0$, we can calculate at what temperature we expect to enter these opaque and viscous regimes. Using the empirical expression for the vapour pressure of Sr from \cite{pucher2025srreferencedata}, and taking $d_0 = \qty{498}{\pico\metre}$ as the van der Waals diameter of a strontium atom, one finds that \mfp{} < $10L$ for $T \gtrsim \qty{570}{\degree}$C. Meanwhile, to reach $\lambda_\text{mf}<d$ would require temperatures in excess of \qty{804}{\degree}C.

Since we operated at oven thermocouple temperatures of \toven{} $\leq$ \qty{500}{\degree}C, it is unlikely that the system achieved a vapour temperature of >\qty{804}{\degree}C, however the addition of the heating laser beam, especially for $\Delta t = \qty{40}{\second}$, could potentially lead to a local Sr vapour temperature of >\qty{500}{\degree}C. While the thermocouple temperatures increased only by $\mathcal{O}(1)$ \qty{}{\degree}C during these pulses, we observed significantly larger increases when leaving the heating laser on continuously for hours; at a baseline \toven{} = \qty{475}{\degree}C from electrical heating only, a nominal \qty{18}{\watt} of heating laser light applied continuously for \qty{260}{\minute} increased the thermocouple temperature to \qty{525}{\degree}C, at which it had settled. We therefore cannot rule out the possibility that a local Sr vapour temperature is being reached that is causing the flow to enter an opaque regime, broadening the lineshape.

The gradual crossover from FMF to other regimes can only be taken as a guideline for our situation where the number density changes along the channel. Moreover, uncertainties in the value of the mean free path arise because the collision cross-section has a dependence on relative velocity (for models more realistic than considering atoms as hard spheres). We are now running simulations of systems that include interatomic collisions, and hope to use these to confirm whether this mechanism could be contributing to the observed transition to a Gaussian profile.

The transition from the theoretical to a Gaussian form has been seen before, for example in \cite{staub2019developing}, wherein the author suggests that it could be a result of the nozzle becoming partially clogged. This would mean a macroscopic amount of Sr is building up in/on the nozzle in a steady-state, and as this is liberated from the nozzle, a smaller fraction of the microchannels' length is used. This acts to reduce the effective length of the microchannel, meaning the aspect ratio, $\beta$, is larger than intended and the extent of the collimation is reduced. A similar counter-argument can be made, however, that such partial clogging would also reduce the effective diameter of the microchannels, which would act to decrease $\beta$ and actually improve the collimation. Other changes to the kinetics at high temperatures, e.g. in the velocity distribution, have also been observed \cite{hill2014zeeman}.

A further alternative mechanism would be for some of the atoms that collide with the channel walls to undergo entirely or partially specular reflection, rather than purely diffuse, as is assumed. This was suggested to be likely by Clausing in his original formulation \cite{clausing1971flow}. The tangential momentum accommodation coefficient (TMAC) is a measure of the extent to which a reflection is diffuse/specular, taking the value 0 for entirely specular reflections, and 1 for entirely diffuse \cite{agrawal2008survey}. A TMAC of zero would therefore mean atoms leave the microchannel at the same angle as they enter it, meaning the nozzle has no collimating effect at all, and the angular distribution out the nozzle would be a pure cosine; this results in a more significant broadening than we see, as shown in orange in Figure\ \ref{fig-gaussian}. For a finite, non-unity TMAC, some fraction of the atoms will be passed through the microchannel after undergoing partially specular reflections, and the angular distribution of atoms out the nozzle would be some intermediate form between the uncollimated cosine and the exactly collimated $j(\theta)$ described in Appendix \ref{sec-app-lineshape-flux} above.





\vspace{10mm}

\twocolumngrid

\bibliography{source_bib}

@Article{feng2024high,
  author    = {Feng, CH and Robert, P and Bouyer, P and Canuel, B and Li, Jianing and Das, Swarup and Kwong, Chang Chi and Wilkowski, David and Prevedelli, M and Bertoldi, A},
  journal   = {Quantum Science and Technology},
  title     = {High flux strontium atom source},
  year      = {2024},
  number    = {2},
  pages     = {025017},
  volume    = {9},
  publisher = {IOP Publishing},
  doi       = {10.1088/2058-9565/ad310b},
}

@Article{hahn2022comparative,
  author    = {Hahn, Rapha{\"e}l and Battard, Thomas and Boucher, Oscar and Picard, Yan J and Lignier, Hans and Comparat, Daniel and Keriel, Nolwenn-Amandine and Lopez, Colin and Oswald, Emanuel and Reveillard, Morgan and Viteau, Matthieu},
  journal   = {Review of Scientific Instruments},
  title     = {Comparative analysis of recirculating and collimating cesium ovens},
  year      = {2022},
  number    = {4},
  volume    = {93},
  publisher = {AIP Publishing},
  pages     = {043302},
  doi       = {10.1063/5.0085838},
}

@article{schioppo2012compact,
  title={A compact and efficient strontium oven for laser-cooling experiments},
  author={Schioppo, Marco and Poli, Nicola and Prevedelli, Marco and Falke, St and Lisdat, Ch and Sterr, Uwe and Tino, Guglielmo Maria},
  journal={Review of Scientific Instruments},
  volume={83},
  number={10},
  year={2012},
  publisher={AIP Publishing},
  pages={103101},
  doi={10.1063/1.4756936},
}

@article{ross1995high,
  title={High temperature metal atom beam sources},
  author={Ross, KJ and Sonntag, B},
  journal={Review of Scientific Instruments},
  volume={66},
  number={9},
  pages={4409--4433},
  year={1995},
  publisher={American Institute of Physics},
  doi={10.1063/1.1145337},
}

@article{yang2015high,
  title={A high flux source of cold strontium atoms},
  author={Yang, Tao and Pandey, Kanhaiya and Pramod, Mysore Srinivas and Leroux, Frederic and Kwong, Chang Chi and Hajiyev, Elnur and Chia, Zhong Yi and Fang, Bess and Wilkowski, David},
  journal={The European Physical Journal D},
  volume={69},
  pages={226},
  year={2015},
  publisher={Springer},
  doi={10.1140/epjd/e2015-60288-y},
}

@article{oxley2016precision,
  title={Precision atomic beam density characterization by diode laser absorption spectroscopy},
  author={Oxley, Paul and Wihbey, Joseph},
  journal={Review of Scientific Instruments},
  volume={87},
  number={9},
  year={2016},
  publisher={AIP Publishing},
  pages={093103},
  doi={10.1063/1.4962025},
}

@article{song2016cost,
  title={A cost-effective high-flux source of cold ytterbium atoms},
  author={Song, Bo and Zou, Yueyang and Zhang, Shanchao and Cho, Chang-woo and Jo, Gyu-Boong},
  journal={Applied Physics B},
  volume={122},
  pages={250},
  year={2016},
  publisher={Springer},
  doi={10.1007/s00340-016-6523-8},
}

@article{lebedev2017note,
  title={Note: Micro-channel array crucible for isotope-resolved laser spectroscopy of high-temperature atomic beams},
  author={Lebedev, Vyacheslav and Bartlett, Joshua H and Malyzhenkov, Alexander and Castro, Alonso},
  journal={Review of Scientific Instruments},
  volume={88},
  number={12},
  year={2017},
  publisher={AIP Publishing},
  pages={126101},
  doi={10.1063/1.5006457},
}

@article{cooper2018collimated,
  title={Collimated dual species oven source and its characterisation via spatially resolved fluorescence spectroscopy},
  author={Cooper, Nathan and Da Ros, Elisa and Nute, Jonathan and Baldolini, Daniele and Jouve, Pierre and Hackerm{\"u}ller, Lucia and Langer, M},
  journal={Journal of Physics D: Applied Physics},
  volume={51},
  number={10},
  pages={105602},
  year={2018},
  publisher={IOP Publishing},
  doi={10.1088/1361-6463/aaa285},
}

@article{nosske2017two,
  title={Two-dimensional magneto-optical trap as a source for cold strontium atoms},
  author={Nosske, Ingo and Couturier, Luc and Hu, Fachao and Tan, Canzhu and Qiao, Chang and Blume, Jan and Jiang, YH and Chen, Peng and Weidem{\"u}ller, Matthias},
  journal={Physical Review A},
  volume={96},
  number={5},
  pages={053415},
  year={2017},
  publisher={APS},
  doi={10.1103/PhysRevA.96.053415},
}

@misc{huckans2018note,
  title={Note on the reflectance of mirrors exposed to a strontium beam}, 
  author={J. Huckans and W. Dubosclard and E. Mar\'{e}chal and O. Gorceix and B. Laburthe-Tolra and M. Robert-de-Saint-Vincent},
  year={2018},
  eprint={1802.08499},
  archivePrefix={arXiv},
  primaryClass={physics.atom-ph},
}

@article{barbiero2020sideband,
  title = {Sideband-Enhanced Cold Atomic Source for Optical Clocks},
  author = {Barbiero, Matteo and Tarallo, Marco G. and Calonico, Davide and Levi, Filippo and Lamporesi, Giacomo and Ferrari, Gabriele},
  journal = {Phys. Rev. Appl.},
  volume = {13},
  issue = {1},
  pages = {014013},
  numpages = {13},
  year = {2020},
  month = {Jan},
  publisher = {American Physical Society},
  doi = {10.1103/PhysRevApplied.13.014013},
}

@article{kobayashi2020demonstration,
  doi = {10.1088/1681-7575/ab9f1f},
  year = {2020},
  month = {nov},
  publisher = {IOP Publishing},
  volume = {57},
  number = {6},
  pages = {065021},
  author = {Kobayashi, Takumi and Akamatsu, Daisuke and Hosaka, Kazumoto and Hisai, Yusuke and Wada, Masato and Inaba, Hajime and Suzuyama, Tomonari and Hong, Feng-Lei and Yasuda, Masami},
  title = {Demonstration of the nearly continuous operation of an $^171$Yb optical lattice clock for half a year},
  journal = {Metrologia},
}

@article{bowden2016adaptable,
  author = {Bowden, William and Gunton, Will and Semczuk, Mariusz and Dare, Kahan and Madison, Kirk W.},
  title = {An adaptable dual species effusive source and Zeeman slower design demonstrated with Rb and Li},
  journal = {Review of Scientific Instruments},
  volume = {87},
  number = {4},
  pages = {043111},
  year = {2016},
  month = {04},
  issn = {0034-6748},
  doi = {10.1063/1.4945567},
}

@article{hsu2022laser,
  author={Hsu, Chung Chuan
  and Larue, R{\'e}my
  and Kwong, Chang Chi
  and Wilkowski, David},
  title={Laser-induced thermal source for cold atoms},
  journal={Scientific Reports},
  year={2022},
  month={Jan},
  day={18},
  volume={12},
  number={1},
  pages={868},
  issn={2045-2322},
  doi={10.1038/s41598-021-04697-4},
}

@article{yasuda2017laser,
  author = {Yasuda, Masami and Tanabe, Takehiko and Kobayashi, Takumi and Akamatsu, Daisuke and Sato, Takumi and Hatakeyama, Atsushi},
  title = {Laser-Controlled Cold Ytterbium Atom Source for Transportable Optical Clocks},
  journal = {Journal of the Physical Society of Japan},
  volume = {86},
  number = {12},
  pages = {125001},
  year = {2017},
  doi = {10.7566/JPSJ.86.125001},
}

@article{kock2016laser,
  author={Kock, Ole and He, Wei and {\'{S}}wierad, Dariusz and Smith, Lyndsie and Hughes, Joshua and Bongs, Kai and Singh, Yeshpal},
  title={Laser controlled atom source for optical clocks},
  journal={Scientific Reports},
  year={2016},
  month={Nov},
  day={18},
  volume={6},
  number={1},
  pages={37321},
  issn={2045-2322},
  doi={10.1038/srep37321},
}

@article{anderson2001loading,
  title = {Loading a vapor-cell magneto-optic trap using light-induced atom desorption},
  author = {Anderson, B. P. and Kasevich, M. A.},
  journal = {Phys. Rev. A},
  volume = {63},
  issue = {2},
  pages = {023404},
  numpages = {6},
  year = {2001},
  month = {Jan},
  publisher = {American Physical Society},
  doi = {10.1103/PhysRevA.63.023404},
}

@article{letellier2023loading,
  author = {Letellier, Hector and Mitchell Galv\~{a}o de Melo, \'{A}lvaro and Dorne, Ana\"{i}s and Kaiser, Robin},
  title = {Loading of a large Yb MOT on the $^1$S$_0$ → $^1$P$_1$ transition},
  journal = {Review of Scientific Instruments},
  volume = {94},
  number = {12},
  pages = {123203},
  year = {2023},
  month = {12},
  issn = {0034-6748},
  doi = {10.1063/5.0169772},
}

@article{senaratne2015effusive,
  author = {Senaratne, Ruwan and Rajagopal, Shankari V. and Geiger, Zachary A. and Fujiwara, Kurt M. and Lebedev, Vyacheslav and Weld, David M.},
  title = {Effusive atomic oven nozzle design using an aligned microcapillary array},
  journal = {Review of Scientific Instruments},
  volume = {86},
  number = {2},
  pages = {023105},
  year = {2015},
  month = {02},
  issn = {0034-6748},
  doi = {10.1063/1.4907401},
}

@article{li2022bicolor,
    author = {Li, Jianing and Lim, Kelvin and Das, Swarup and Zanon-Willette, Thomas and Feng, Chen-Hao and Robert, Paul and Bertoldi, Andrea and Bouyer, Philippe and Kwong, Chang Chi and Lan, Shau-Yu and Wilkowski, David},
    title = {Bi-color atomic beam slower and magnetic field compensation for ultracold gases},
    journal = {AVS Quantum Science},
    volume = {4},
    number = {4},
    pages = {046801},
    year = {2022},
    month = {12},
    issn = {2639-0213},
    doi = {10.1116/5.0126745},
}

@article{agrawal2008survey,
    author = {Agrawal, Amit and Prabhu, S. V.},
    title = {Survey on measurement of tangential momentum accommodation coefficient},
    journal = {Journal of Vacuum Science \& Technology A},
    volume = {26},
    number = {4},
    pages = {634-645},
    year = {2008},
    month = {06},
    issn = {0734-2101},
    doi = {10.1116/1.2943641},
}

@Inbook{pauly2000fundamentals,
    author="Pauly, Hans",
    title="{Fundamentals of Kinetic Gas Theory}",
    bookTitle="Atom, Molecule, and Cluster Beams I: Basic Theory, Production and Detection of Thermal Energy Beams",
    year="2000",
    publisher="Springer Berlin Heidelberg",
    pages="35-76",
    isbn="978-3-662-04213-7",
    doi="10.1007/978-3-662-04213-7_2",
    url="https://doi.org/10.1007/978-3-662-04213-7_2"}

@article{clausing1971flow,
    author = {Clausing, P.},
    title = {The Flow of Highly Rarefied Gases through Tubes of Arbitrary Length},
    journal = {Journal of Vacuum Science and Technology},
    volume = {8},
    number = {5},
    pages = {636-646},
    year = {1971},
    month = {09},
    issn = {0022-5355},
    doi = {10.1116/1.1316379},
    url = {https://doi.org/10.1116/1.1316379},
}

@mastersthesis{staub2019developing,
    author = {Staub, Etienne},
    title = {Developing a High-Flux Atomic Beam Source for Experiments with Ultracold Strontium Quantum Gases},
    school = {LMU, M\"{u}nchen},
    year = {2019},
    url = {https://ultracold.sr/publications/thesis_etienne_staub.pdf}
}

@article{gao2021optically,
    author = {Gao, S. and Hughes, W. J. and Lucas, D. M. and Ballance, T. G. and Goodwin, J. F.},
    title = {An optically heated atomic source for compact ion trap vacuum systems},
    journal = {Review of Scientific Instruments},
    volume = {92},
    number = {3},
    pages = {033205},
    year = {2021},
    month = {03},
    issn = {0034-6748},
    doi = {10.1063/5.0038162},
    url = {https://doi.org/10.1063/5.0038162},
}

@misc{pucher2025srreferencedata,
      title={$^{88}${Sr} Reference Data}, 
      author={Sebastian Pucher and Sofus Laguna Kristensen and Ronen M. Kroeze},
      year={2025},
      eprint={2507.10487},
      archivePrefix={arXiv},
      primaryClass={physics.atom-ph},
}

@article{nist2010srdata,
    author = {Sansonetti, J. E. and Nave, G.},
    title = {Wavelengths, Transition Probabilities, and Energy Levels for the Spectrum of Neutral Strontium (SrI)},
    journal = {Journal of Physical and Chemical Reference Data},
    volume = {39},
    number = {3},
    pages = {033103},
    year = {2010},
    month = {08},
    issn = {0047-2689},
    doi = {10.1063/1.3449176},
    url = {https://doi.org/10.1063/1.3449176},
}

@article{ye2013production,
  title = {Production of very-high-$n$ strontium Rydberg atoms},
  author = {Ye, S. and Zhang, X. and Killian, T. C. and Dunning, F. B. and Hiller, M. and Yoshida, S. and Nagele, S. and Burgd\"orfer, J.},
  journal = {Phys. Rev. A},
  volume = {88},
  issue = {4},
  pages = {043430},
  numpages = {10},
  year = {2013},
  month = {Oct},
  publisher = {American Physical Society},
  doi = {10.1103/PhysRevA.88.043430},
  url = {https://link.aps.org/doi/10.1103/PhysRevA.88.043430}
}

@Article{bowden2019realize,
    author={Bowden, William and Hobson, Richard and Hill, Ian R. and Vianello, Alvise and Schioppo, Marco and Silva, Alissa and Margolis, Helen S. and Baird, Patrick E. G. and Gill, Patrick},
    title={A pyramid MOT with integrated optical cavities as a cold atom platform for an optical lattice clock},
    journal={Scientific Reports},
    year={2019},
    month={Aug},
    day={12},
    volume={9},
    number={1},
    pages={11704},
    issn={2045-2322},
    doi={10.1038/s41598-019-48168-3},
    url={https://doi.org/10.1038/s41598-019-48168-3}
}

@article{campbell2017some,
    author = {S. L. Campbell  and R. B. Hutson  and G. E. Marti  and A. Goban  and N. Darkwah Oppong  and R. L. McNally  and L. Sonderhouse  and J. M. Robinson  and W. Zhang  and B. J. Bloom  and J. Ye },
    title = {A Fermi-degenerate three-dimensional optical lattice clock},
    journal = {Science},
    volume = {358},
    number = {6359},
    pages = {90-94},
    year = {2017},
    doi = {10.1126/science.aam5538},
    URL = {https://www.science.org/doi/abs/10.1126/science.aam5538}
}

@article{thekkeppatt2025measurement,
  title = {Measurement of the $g$ Factor of Ground-State $^{87}\mathrm{Sr}$ at the Parts-per-Million Level Using Co-Trapped Ultracold Atoms},
  author = {Thekkeppatt, Premjith and Digvijay and Urech, Alexander and Schreck, Florian and van Druten, Klaasjan},
  journal = {Phys. Rev. Lett.},
  volume = {135},
  issue = {19},
  pages = {193001},
  numpages = {8},
  year = {2025},
  month = {Nov},
  publisher = {American Physical Society},
  doi = {10.1103/cjks-9hlp},
  url = {https://link.aps.org/doi/10.1103/cjks-9hlp}
}

@misc{graham2017midbandgravitationalwavedetection,
      title={Mid-band gravitational wave detection with precision atomic sensors}, 
      author={Peter W. Graham and Jason M. Hogan and Mark A. Kasevich and Surjeet Rajendran and Roger W. Romani},
      year={2017},
      eprint={1711.02225},
      archivePrefix={arXiv},
      primaryClass={astro-ph.IM},
}

@article{escudero2021steady,
  title = {Steady-state magneto-optical trap of fermionic strontium on a narrow-line transition},
  author = {Escudero, Rodrigo Gonz\'alez and Chen, Chun-Chia and Bennetts, Shayne and Pasquiou, Benjamin and Schreck, Florian},
  journal = {Phys. Rev. Res.},
  volume = {3},
  issue = {3},
  pages = {033159},
  numpages = {11},
  year = {2021},
  month = {Aug},
  publisher = {American Physical Society},
  doi = {10.1103/PhysRevResearch.3.033159},
  url = {https://link.aps.org/doi/10.1103/PhysRevResearch.3.033159}
}

@Article{Alonso2022,
    title={Cold atoms in space: community workshop summary and proposed road-map},
    journal={EPJ Quantum Technology},
    year={2022},
    month={Nov},
    day={20},
    volume={9},
    number={1},
    pages={30},
    issn={2196-0763},
    doi={10.1140/epjqt/s40507-022-00147-w},
    url={https://doi.org/10.1140/epjqt/s40507-022-00147-w},
    author={Alonso, Iv{\'a}n and Alpigiani, Cristiano and Altschul, Brett and Ara{\'u}jo, Henrique and Arduini, Gianluigi and Arlt, Jan and Badurina, Leonardo and Bala{\v{z}}, Antun and Bandarupally, Satvika and Barish, Barry C. and Barone, Michele and Barsanti, Michele and Bass, Steven and Bassi, Angelo and Battelier, Baptiste and Baynham, Charles F. A. and Beaufils, Quentin and Beli{\'{c}}, Aleksandar and Berg{\'e}, Joel and Bernabeu, Jose and others}}

@article{stellmer2013production,
  title = {Production of quantum-degenerate strontium gases},
  author = {Stellmer, Simon and Grimm, Rudolf and Schreck, Florian},
  journal = {Phys. Rev. A},
  volume = {87},
  issue = {1},
  pages = {013611},
  numpages = {16},
  year = {2013},
  month = {Jan},
  publisher = {American Physical Society},
  doi = {10.1103/PhysRevA.87.013611},
  url = {https://link.aps.org/doi/10.1103/PhysRevA.87.013611}
}

@misc{NIST-ASD,
    author = {Kramida, A. and Ralchenko, Yu. and Reader, J. and {NIST ASD Team}},
    year = {2024},
    title = {{NIST Atomic Spectra Database (version 5.12)}},
    howpublished = {\url{https://physics.nist.gov/asd}},
    publisher = {National Institute of Standards and Technology},
    doi = {10.18434/T4W30F},
    note = {Accessed: 2026-01-30}
}

@Article{fartmann2025ramsey,
    author={Fartmann, Oliver and Jutisz, Martin and Mahdian, Amir and Schkolnik, Vladimir and Tietje, Ingmari C. and Zimmermann, Conrad and Krutzik, Markus},
    title={Ramsey-Bord{\'e} atom interferometry with a thermal strontium beam for a compact optical clock},
    journal={EPJ Quantum Technology},
    year={2025},
    month={Mar},
    day={03},
    volume={12},
    number={1},
    pages={31},
    issn={2196-0763},
    doi={10.1140/epjqt/s40507-025-00332-7}
}

@article{lucas1973production,
    title = {The production of intense atomic beams},
    journal = {Vacuum},
    volume = {23},
    number = {11},
    pages = {395-402},
    year = {1973},
    issn = {0042-207X},
    doi = {10.1016/0042-207X(73)92529-3},
    author = {CB Lucas}
}

@book{lucas2013atomic,
  title={Atomic and molecular beams: production and collimation},
  author={Lucas, Cyril Bernard},
  year={2013},
  publisher={CRC press}
}

@article{wouters2016design,
    author = {Wouters, Steinar H. W. and ten Haaf, Gijs and Mutsaers, Peter H. A. and Vredenbregt, Edgar J. D.},
    title = {Design and experimental validation of a compact collimated Knudsen source},
    journal = {Review of Scientific Instruments},
    volume = {87},
    number = {8},
    pages = {083305},
    year = {2016},
    month = {08},
    issn = {0034-6748},
    doi = {10.1063/1.4960997},
}

@article{olander1970molecular,
    author = {Olander, Donald R. and Kruger, Valerie},
    title = {Molecular Beam Sources Fabricated from Multichannel Arrays. III. The Exit Density Problem},
    journal = {Journal of Applied Physics},
    volume = {41},
    number = {7},
    pages = {2769-2776},
    year = {1970},
    month = {06},
    issn = {0021-8979},
    doi = {10.1063/1.1659313},
}

@article{li2020robust,
  title = {Robust characterization of microfabricated atomic beams on a six-month time scale},
  author = {Li, Chao and Wei, Bochao and Chai, Xiao and Yang, Jeremy and Daruwalla, Anosh and Ayazi, Farrokh and Raman, C.},
  journal = {Phys. Rev. Res.},
  volume = {2},
  issue = {2},
  pages = {023239},
  numpages = {14},
  year = {2020},
  month = {May},
  publisher = {American Physical Society},
  doi = {10.1103/PhysRevResearch.2.023239},
  url = {https://link.aps.org/doi/10.1103/PhysRevResearch.2.023239}
}

@article{hill2014zeeman,
    doi = {10.1088/0953-4075/47/7/075006},
    year = {2014},
    month = {mar},
    publisher = {IOP Publishing},
    volume = {47},
    number = {7},
    pages = {075006},
    author = {Hill, Ian R and Ovchinnikov, Yuri B and Bridge, Elizabeth M and Curtis, E Anne and Gill, Patrick},
    title = {Zeeman slowers for strontium based on permanent magnets},
    journal = {Journal of Physics B: Atomic, Molecular and Optical Physics}
}

@book{scoles1988atomic,
  title={Atomic and Molecular Beam Methods},
  editor={Scoles, Giacinto and Bassi, Davide and Buck, Udo and Lain{\'e}, Derek},
  volume={1},
  year={1988},
  publisher={Oxford University Press}
}

@book{ramsey1956molecular,
  title={Molecular Beams},
  author={Ramsey, Norman},
  volume={20},
  year={1956},
  publisher={Oxford University Press}
}

@misc{baynham2025prototype,
      title={A Prototype Atom Interferometer to Detect Dark Matter and Gravitational Waves}, 
      year={2025},
      eprint={2504.09158},
      archivePrefix={arXiv},
      primaryClass={hep-ex},
      author={C. F. A. Baynham and R. Hobson and O. Buchmueller and D. Evans and L. Hawkins and L. Iannizzotto-Venezze and A. Josset and D. Lee and E. Pasatembou and B. E. Sauer and M. R. Tarbutt and T. Walker and O. Ennis and U. Chauhan and A. Brzakalik and S. Dey and S. Hedges and B. Stray and M. Langlois and K. Bongs and others}}

@article{tschersich1998formation,
    author = {Tschersich, K. G. and von Bonin, V.},
    title = {Formation of an atomic hydrogen beam by a hot capillary},
    journal = {Journal of Applied Physics},
    volume = {84},
    number = {8},
    pages = {4065-4070},
    year = {1998},
    month = {10},
    issn = {0021-8979},
    doi = {10.1063/1.368619},
}

@misc{okamoto2026direct,
      title={Direct loading of a {Sr} magneto-optical trap from a thermal atomic beam}, 
      author={Naohiro Okamoto and Takumi Sato and Takatoshi Aoki and Yoshio Torii},
      year={2026},
      eprint={2509.13764},
      archivePrefix={arXiv},
      primaryClass={physics.atom-ph},
}

@book{footbook,
    author = {Foot, C. J.},
    title = {Atomic Physics},
    publisher = {Oxford University Press},
    year = {2004},
    month = {11},
    isbn = {9780198506959},
    doi = {10.1093/oso/9780198506959.001.0001},
    url = {https://doi.org/10.1093/oso/9780198506959.001.0001},
}

@misc{hifais,
  title = {A High-Flux Source of Cold Strontium with a Loading Rate of {$4 \times 10^{10}$} atoms/s for Open Release},
  author = {Walker, Thomas and Marchant, Anna L. and Bentine, Elliot and Buchmueller, Oliver and Clarke, Katherine and Foot, Christopher and Hawkins, Leonie and Hughes, Kenneth M. and Hussain, Kamran and Iannizzotto-Venezze, Ludovico and Josset, Alice and Labiad, Hamza and Lee, Dillen and Thornton-Sparkes, Timothy C. and Valenzuela, Tristan and van der Grinten, Maurits and Vick, Andrew and Bason, Mark G. and Baynham, Charles F. A. and Hobson, Richard},
  year = {2026},
  note = {\textit{In preparation}}
}
\end{document}